\documentclass[preprints,article,accept,pdftex,moreauthors]{Definitions/mdpi} 

\firstpage{1} 
\pubvolume{1}
\issuenum{1}
\articlenumber{0}
\pubyear{2022}
\copyrightyear{2022}
\datereceived{12 October 2022} 
\dateaccepted{10 November 2022} 
\datepublished{24 November 2022} 
\hreflink{https://doi.org/10.3390/galaxies10060107} % If needed use \linebreak
\pdfoutput=1

\newcommand{\suml}{\sum\limits}

\Title{Relativistic Signatures of Flux Eruption Events Near Black Holes}

\TitleCitation{Relativistic Signatures of Flux Eruption Events Near Black Holes}

\Author{Zachary Gelles $^{1,2,3,}$*\orcidA{}, Koushik Chatterjee $^{1,2,}*$\orcidB{}, Michael Johnson $^{1,2}$\orcidC{}, Bart Ripperda $^{4,5}$\orcidD{} and Matthew Liska $^{1,6,\dagger}$}

\AuthorNames{Zachary Gelles, Koushik Chatterjee, Michael Johnson, Bart Ripperda and Matthew Liska}

\AuthorCitation{Gelles, Z.; Chatterjee, K.; Johnson, M.; Ripperda, B.; Liska, M.}
\address{%
$^{1}$ \quad Center for Astrophysics, Harvard \& Smithsonian, 60 Garden Street, Cambridge, MA 02138, USA\\
$^{2}$ \quad Black Hole Initiative, Harvard University, 20 Garden Street, Cambridge, MA 02138, USA\\
$^{3}$ \quad Department of Physics, Princeton University, Princeton, NJ 08544, USA\\
$^{4}$ \quad Department of Astrophysical Sciences, Peyton Hall, Princeton University, Princeton, NJ 08544, USA\\
$^{5}$ \quad Center for Computational Astrophysics, Flatiron Institute, 162 Fifth Avenue, New York, NY 10010, USA\\
$^{6}$ \quad Institute for Theory and Computation (ITC), Harvard University, 60 Garden Street, Cambridge, MA 02138, USA
}

\corres{Correspondence: zgelles@princeton.edu (Z.G.), koushik.chatterjee@cfa.harvard.edu (K.C.), mjohnson@cfa.harvard.edu (M.J.), bripperda@princeton.edu (B.P.), matthew.liska@cfa.harvard.edu (M.L.)}

\firstnote{John Harvard Distinguished Science and ITC Fellow} 

\abstract{Images of supermassive black holes produced using very long baseline interferometry provide a pathway to directly observing effects of a highly curved spacetime, such as a bright ``photon ring'' that arises from strongly lensed emission. In addition, the emission near supermassive black holes is highly variable, with bright high-energy flares regularly observed. We demonstrate that intrinsic variability can introduce prominent associated changes in the relative brightness of the photon ring. We analyze both semianalytic toy models and GRMHD simulations with magnetic flux eruption events, showing that they each exhibit a characteristic ``loop'' in the space of relative photon ring brightness versus total flux density. For black holes viewed at high inclination, the relative photon ring brightness can change by an order of magnitude, even with variations in total flux density that are comparatively mild. We show that gravitational lensing, Doppler boosting, and magnetic field structure all significantly affect this feature, and we discuss the prospects for observing it in observations of M87$^*$ and Sgr~A$^*$ with the next-generation Event Horizon Telescope.}


\begin{document}

\section{Introduction}\label{sec:intro}

The Event Horizon Telescope released the first images of M87$^*$ in 2019 and the first images of Sgr~A$^*$ in 2022 \citep{EHT_1,EHT_sgr_1}, enabling new measurements of black hole accretion flow properties directly from VLBI data and event-horizon-scale images. A theorized component of all black hole images is the photon ring: a thin annulus of light composed of photons traveling on nearly-bound geodesics. This photon ring splits into a series of self-similar subrings, each of which reflects a different degree of light-bending around the hole \citep[see, e.g.,][]{Luminet_1979,deVries_2000,Takahashi_2004,Beckwith_2005,Johannsen_Psaltis_2010,Gralla_2019,Johnson_2020}.

In particular, each subring is labeled by an index $n$,  which counts the number of half-orbits that a photon completes on its trajectory from emitter to observer. In the optically thin limit, each successive subring has similar brightness but is exponentially demagnified, with the demagnification related to Lyapunov exponents that are governed by the properties of unstable spherical orbits of null geodesics \citep[see, e.g.][]{Darwin_1959,Luminet_1979,Ohanian_1987,Johnson_2020}. 

One consequence of the demagnification is that the direct image $(n=0)$ of a black hole tends to be the dominant source of observed flux, with indirect images $(n\geq 1)$ appearing exponentially suppressed. Hence, an important quantity is the Photon ring Flux Ratio (PFR), defined as the fraction of total flux contained in a particular subimage. We focus on the $n=1$ ratio, which we denote as $f_1$:\begin{align}
    f_1&\equiv\frac{F_1}{F_{\rm tot}}.
\end{align}
This quantity depends both on the spacetime (which entirely determines the relative demagnification of the subring), the emission geometry, and the magnetic field structure. Moreover, it can potentially be measured with the next-generation Event Horizon Telescope (ngEHT) by using modeling methods that isolate the contribution of the photon ring \citep[e.g.,][]{Broderick_2020,Tiede_2022}.

\begin{figure*}
\centering
    \includegraphics[width=\textwidth]{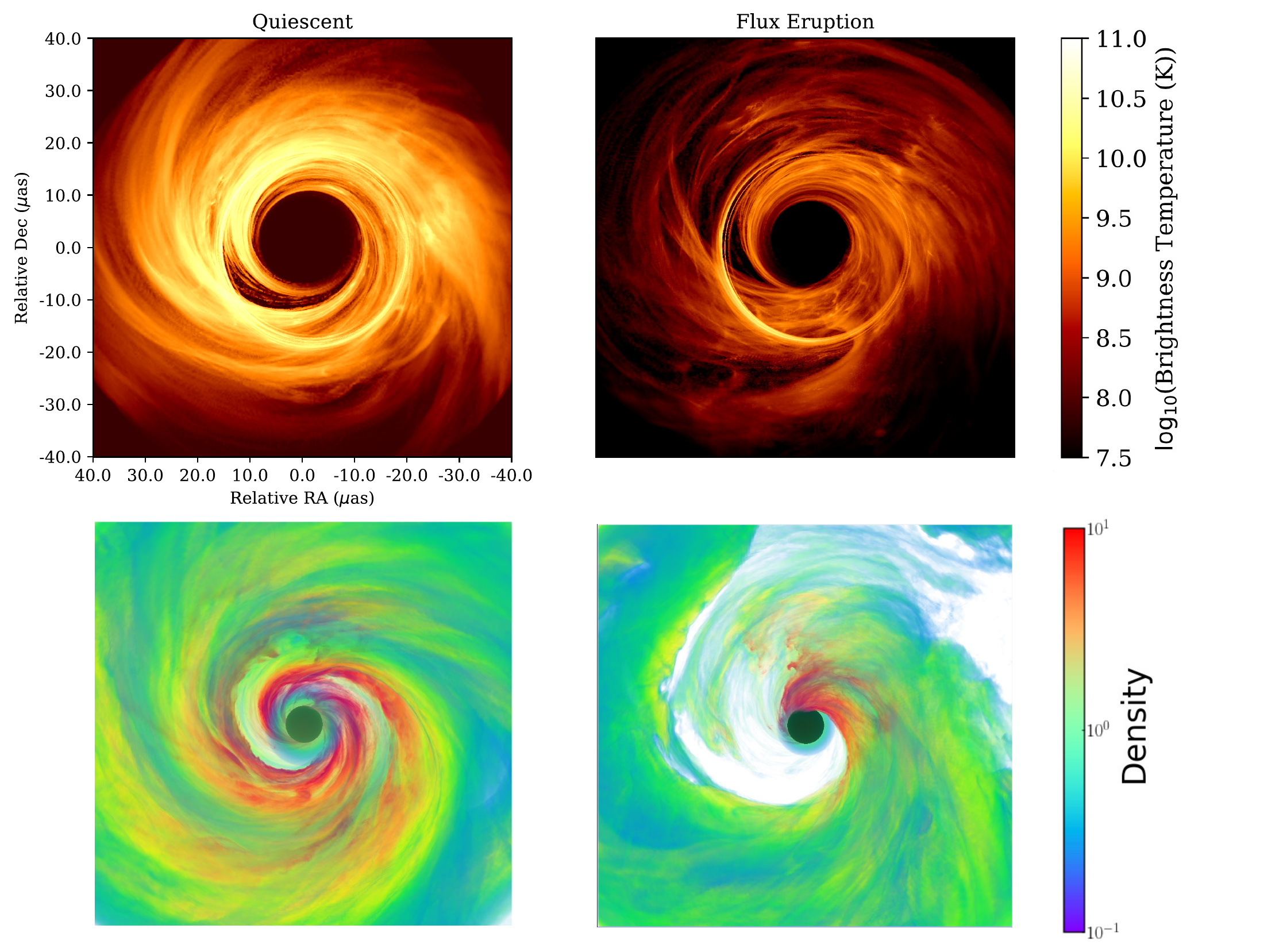}
	\caption{Magnetic flux eruptions can remove over half of the disk from near the black hole, significantly changing the resultant horizon-scale image. Top: Snapshots of simulation in quiescent state (left) and flux eruption event (right), ray-traced with equal mass scale units at a viewing inclination of $17^\circ$. Bottom: 3D rendering of gas density (in GRMHD code units) for the quiescent state (left) and flux eruption event (right) within the inner $15$M, viewed at $17^\circ$. The white region in the density plot shows the evacuation of the disk and the formation of a low density magnetospheric region near the black hole.}
    \label{fig:comparefig}
\end{figure*}

General relativistic magneto-hydrodynamic (GRMHD) simulations and general relativistic radiative transfer (GRRT) are important tools for connecting observations of the photon ring to the underlying plasma and emission physics of the accretion disk \citep[e.g.,][]{Porth_2019,Gold_2020}. GRMHD simulations used to interpret the 2017 EHT observations favor strongly magnetized gas accreting onto Sgr A$^*$ and M87$^*$ \citep{EHT_5,EHT_sgr_5}. In the magnetically arrested disk \citep[MAD,][]{Narayan_2003,Igumenshchev_2003} limit, the magnetic field near the BH becomes strong enough to vertically squeeze the accreting gas. These fields ultimately undergo magnetic reconnection, allowing a bundle of vertical fields to escape from the vicinity of the BH, and in the process, eject out a large portion of the disk \citep[e.g.][]{Ripperda_2022}. Observable signatures of such MAD system behavior in Sgr A* were recently reported by \citet{Wielgus:2022b}. These ``magnetic flux eruption events'' exhibit large gas temperatures, strong vertical fields and occur quasi-periodically in the MAD state \citep[e.g.][]{Narayan_2022}, and hence, are a prime candidate for the origin of high-energy flux eruptions, e.g., in Sgr A$^*$ and M87$^*$. An example of such a flux eruption is depicted in Figure~\ref{fig:comparefig}, and a more detailed description of the accretion flow's response to these events is outlined in Appendix~\ref{sec:grmhdapp}.

During these eruptions, the flux density in the image can drop by a factor of ${\sim}$10 (see Jia et al., in prep for more details about the lightcurve). Observations of Sagittarius A* indicate a flux density drop of a factor of 2 at millimeter wavelengths following a high energy flaring event \citep[][]{EHT_sgr_2,Wielgus:2022a}. Furthermore, simulations indicate that individual eruptions may last for ${\sim}100-300 M$,\footnote{We use natural units, i.e., $G=c=1$, and the length and time units are both defined only in terms of the black hole mass $M$.} with the flux bundle eventually dissipating in the disk. This timescale corresponds to a few weeks to months for M87$^*$ or an ${\sim}$hour for Sgr~A$^*$, and is reasonably consistent with the analysis of Sgr A* X-ray flares population \citep[][]{Haggard:2019}. Such a long evolution period allows us to possibly capture eruption events with instruments such as the ngEHT and to resolve the flaring state of a SMBH. The presence of strong vertical fields and the ejection of gas present an exceptional opportunity to probe the detailed structure of the flux eruption through its observable features in polarization and variability \citep[e.g.,][]{GRAVITY:2021, Gelles_2021b}. 

In this work, we explore how the dynamics of flux eruptions are manifest in image morphology, with specific attention to the underlying factors that directly control the relative brightness of the photon ring during these events. In Section~\ref{sec:toymodel}, we introduce a toy model to simplify the depiction of a flux eruption event, allowing us to separate the individual influences of Doppler boosting and gravitational lensing on the photon ring's appearance. In Section~\ref{sec:grmhdsec}, we repeat a similar analysis on a set of GRMHD data, showing that the results of the toy model are recovered provided that the emission profile is not changed. In Section~\ref{sec:magnetism}, we show that the introduction of a magnetic field (with a thermal emission profile) dramatically alters the photon ring brightness. Finally, in Section~\ref{sec:discussion}, we discuss our results in the context of observable targets for future EHT and ngEHT science.

\section{Toy Model}\label{sec:toymodel}

In this section, we develop a toy model to illuminate the effects of Doppler boosting on the PFR during flux eruption events. 
\subsection{Description of the Model}
During a flux eruption event, a substantial portion of the high-density material in the disk is ejected, leaving a low-density magnetosphere (with an equatorial current sheet) in the inner region. Our toy model therefore consists of a half-wedge (i.e., a wedge that spans $180^\circ$ in $\phi$) surrounding a Kerr black hole, with emission extending down to the horizon. In practice, one could experiment with smaller wedges that span a narrower range of $\phi$ values, but we restrict our focus to the half-wedge for simplicity. The emission from the wedge represents the near-horizon sub-millimeter emission seen during a flux eruption event.

From here, we build off of the semianalytic models of \citet{Gold_2020}, modifying the prescriptions so that emission is confined to exactly one half of the spacetime. Among these semianalytic test models, we adapt ``Test 5," which entails a simple, isotropic emissivity function (i.e., independent of magnetic field direction) and a thin scale height for the disk:
\begin{align}
    n(\vec{r})&\propto \exp\left\{-\frac{1}{2}\left[\left(\frac{r}{10}\right)^2+\left(\frac{100}{3}\cos\theta\right)^2\right]\right\},\\
   \nonumber j_\nu(\vec{r})&\propto n(\vec{r}),\\
 \nonumber   \alpha_\nu(\vec{r})&=10^6\left(\frac{\nu}{230\,{\rm GHz}}\right)^{-2.5}\times j_\nu(\vec{r}).
\end{align}
Here, $n$ is the number density, $j_\nu$ is the emissivity, and $\alpha_\nu$ is the absorptivity, with $\theta$ and $r$ taking on their Boyer-Lindquist coordinate values. In addition to cutting out half of the emitting region, we also modify the original prescription of \citet{Gold_2020} so that distance of the black hole matches that of M87$^*$, the spin of the black hole is $a=+0.94$, and the overall density is rescaled so $F_{230}\sim 0.5$ Jy for one of the snapshots.

\subsection{Photon Ring Flux Ratios}
During a flux eruption event, there is a disruption of the approximate axisymmetry in the accretion structure. One expects the reconnection layer powering the eruption to emit strongly in the X-ray, leaving the cool gas in the disk to emit in the sub-millimeter and radio. Hence, we expect the PFR to depend strongly on the position of the observer relative to the flux eruption itself. This orientation is encoded through both the observer's polar inclination angle ($\theta_{\rm cam}$), which is measured from the black hole's spin axis, as well as the observer's azimuthal angle ($\phi_{\rm cam}$). We orient the azimuthal coordinate so that $\phi_{\rm cam}=0$ when the region of highest density (and hence highest emissivity) is positioned directly in front of the hole. In a physical black hole, such a scenario would take place when the flux tube is positioned directly \emph{behind} the black hole.

We ray-trace the toy model for a range of values of $\phi_{\rm cam}$ as a proxy for tracking the flux eruption over time, as increasing $\phi_{\rm cam}$ is equivalent to rotating the accretion disk with respect to the observer. We ray-trace the disk for both a ``static'' fluid rotation profile ($u_\phi = 0$) as well as an asymptotically Keplerian rotation profile ($u_\phi\to (r\sin\theta)^{1/2}$ as $r\to \infty$), corresponding to $\ell_0=0$ and $\ell_0=1$ in \citet{Gold_2020}, respectively. This allows us to isolate the effects of Doppler boosting, which is relevant only when objects are in motion. We further repeat this procedure for $\theta_{\rm cam}=17^\circ$ and $\theta_{\rm cam}=80^\circ$, representing both the low-inclination and high-inclination limits of the viewing geometry.

The ray-tracing is performed using the adaptive sampling scheme of \texttt{ipole} \citep{ipole_2018,Gelles_2021a}. The field of view is taken to be $160\,\mu$as with an effective pixel size of $\sim 0.15\,\mu$as, sufficient to fully resolve the $n=1$ subring. Each image is decomposed into its individual subrings using the procedure described by \citet{Gelles_2021a}. In this scheme, the intensity of a pixel in the $n^{\rm th}$ subimage is computed by performing radiative transfer along the corresponding geodesic's $n^{\rm th}$ pass around the black hole. 

\begin{figure*}
\centering
    \includegraphics[width=0.68\textwidth]{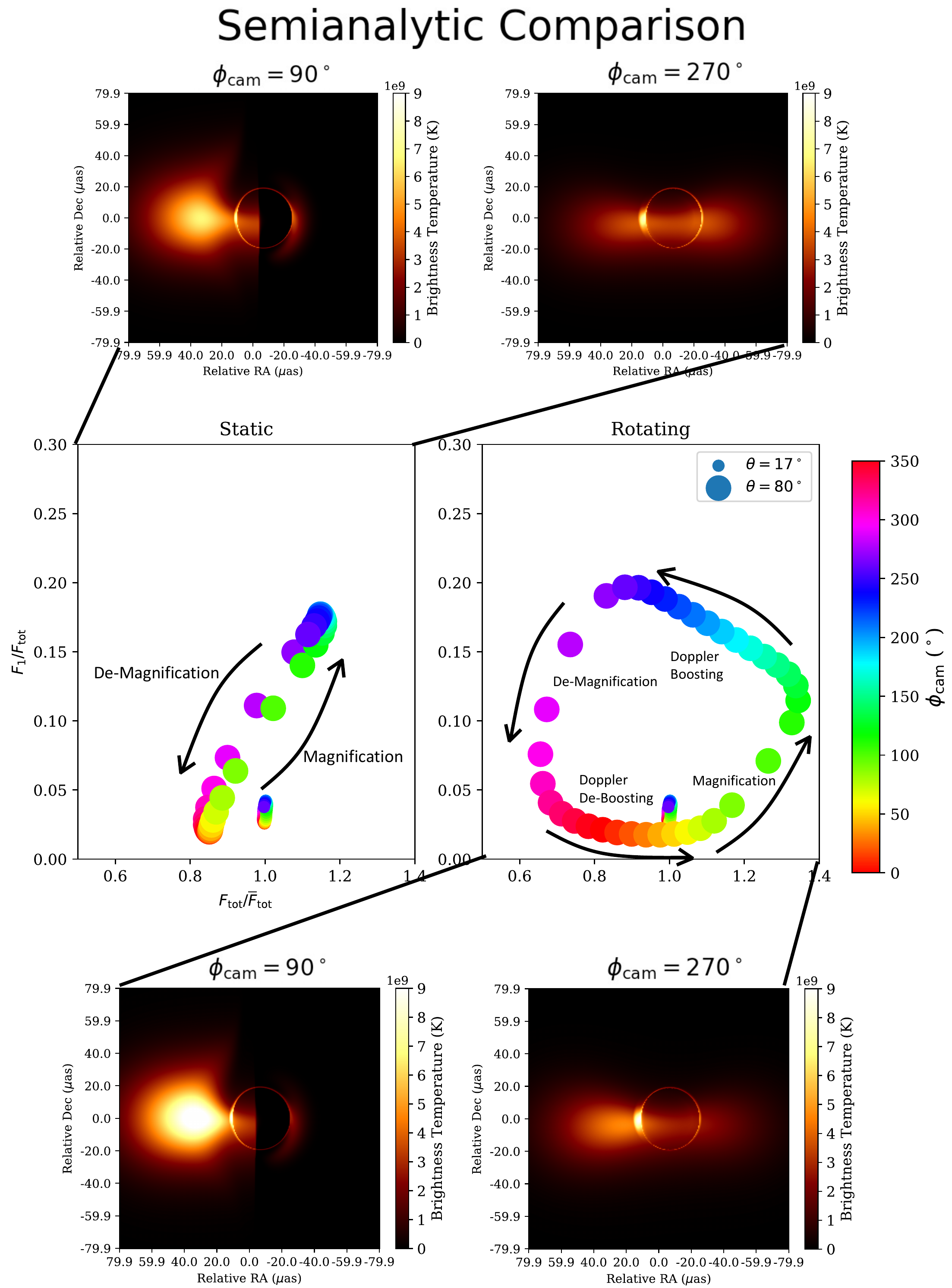}
    \includegraphics[width=0.27\textwidth]{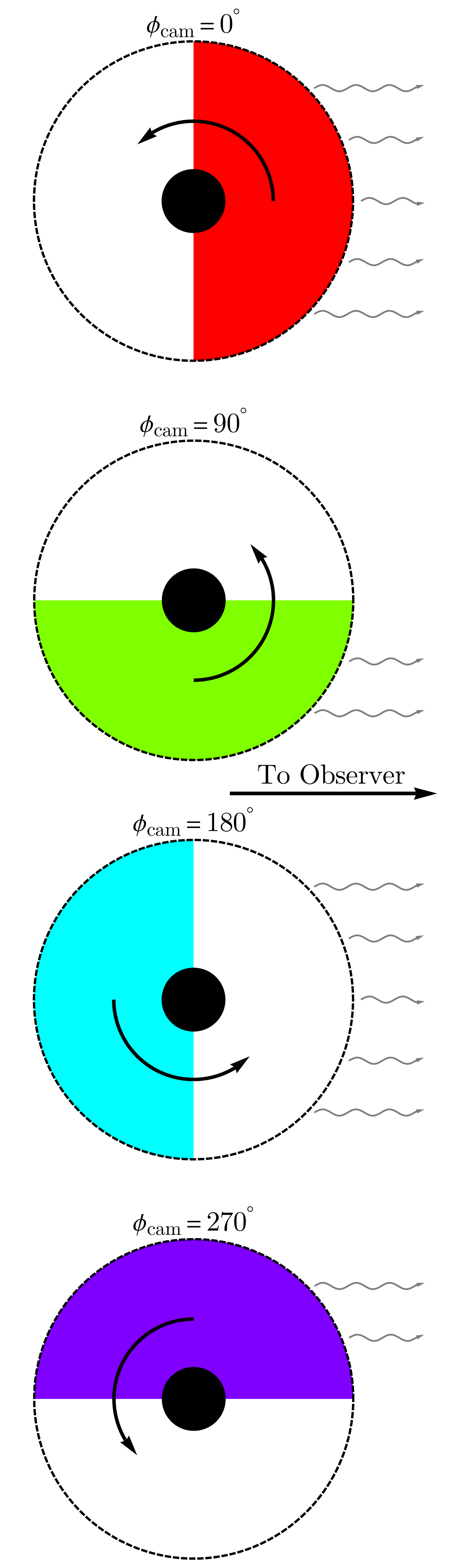}
	\caption{PFR in half-disk toy model, along with a schematic depicting the disk orientation on the right. Snapshots are for ``static" disk (top) and rotating disk (bottom), both with $\theta_{\rm cam}=80^\circ$. Both curves show a steep change in the PFR with $\phi_{\rm cam}$ because of gravitational lensing, but the addition of rotation introduces Doppler effects that modulate the relative flux, spreading the curve horizontally. The Doppler effects are also seen to a small degree for the ``static'' case because of the angular velocity of the zero-angular momentum frame.}
    \label{fig:semifig}
\end{figure*}

In Figure~\ref{fig:semifig}, we plot the ratio $f_1$ for each of these configurations and rotation profiles as a function of the total image flux. We normalize the total image flux $F_{\rm tot}$ with a mean flux $\overline{F}_{\rm tot}$ taken over all values of $\phi_{\rm cam}$, which effectively represents a time-averaged flux. Since we may be able to measure the ratio $f_1$ along with the total compact flux using the ngEHT, we choose to plot $f_1$ against $F_{\rm tot}/\overline{F}_{\rm tot}$ for each camera azimuthal position $\phi_{\rm cam}$. The flux eruption induces a strong correlation between the total image flux and the fractional flux contained in the subring. As the half-disk orbits around the black hole, the PFR changes dramatically, more than doubling over the course of one revolution in the edge-on limit.

The specific shape of the curves traced out in Figure~\ref{fig:semifig} is due to a confluence of numerous factors. In the low inclination limit ($\theta_{\rm cam}=17^\circ$), the PFR remains relatively constant over azimuth, regardless of whether or not the disk is rotating. This is because the fluid velocity is confined to the midplane, so with a low-inclination viewing geometry, the half-disk cannot acquire a large velocity tangent to null geodesics that reach the observer.

The specific value of the PFR is consistent with theoretical predictions as well. For an $a=+0.94$ Kerr black hole viewed at $\theta_{\rm cam}=17^\circ$, the Lyapunov exponent ranges from $\gamma= 2.36$ to $\gamma= 2.76$, depending on the azimuthal screen coordinate. Following \citet{Johnson_2020}, in the asymptotic limit of large $n$ (as well as optical transparency and axisymmetry), one expects 

\begin{align}
    \frac{F_{n+1}}{F_n} &\sim e^{-\gamma}\\
    \nonumber \Longrightarrow \frac{F_1}{F_{\rm tot}} &= \frac{F_1}{F_0}\frac{1}{1+\frac{F_1}{F_0}+...}\\
    \nonumber &\sim e^{-\gamma}\frac{1}{\suml_{n=0}^\infty e^{-\gamma n}}\\
    \nonumber &= e^{-\gamma}-e^{-2\gamma},
\end{align}
which ranges from $0.059$ to $0.085$ for this particular black hole. This range falls not far from the PFR of the low inclination snapshots in Figure~\ref{fig:semifig}. 

On the other hand, in the high-inclination case ($\theta_{\rm cam}=80^\circ$), the brightness of the photon ring depends heavily on the location of the observer with respect to the flux eruption. In the case of a static rotation profile (Figure~\ref{fig:semifig}; left panel), the PFR is primarily determined by the relative magnification from gravitational lensing. As the half-disk passes behind the black hole, more of the received flux is bent around the hole, leading to a magnification of the $n=1$ image and a resultant brightening of the photon ring. Indeed, the PFR is largest when $\phi_{\rm cam}\sim 180^\circ$ and the emission wedge is located behind the black hole. The phases of increasing magnification and demagnification are demarcated with arrows in Figure~\ref{fig:semifig}. 

However, when the disk is rotating (Figure~\ref{fig:semifig}; right panel), Doppler effects become important. The maximum and minimum PFR's still occur when $\phi_{\rm cam}\sim180^\circ$ and $\phi_{\rm cam}\sim 0^\circ$ respectively, as the gravitational lensing is identical to the non-rotating case. However, Doppler boosting stretches the PFR curve out horizontally, as the recessional speed of the eruption is largest when $\phi_{\rm cam}\sim 90^\circ$ and $\phi_{\rm cam}\sim 270^\circ$. Indeed, in the bottom left snapshot of Figure~\ref{fig:semifig}, one can see the Doppler boosted direct image of the flux eruption on the left, leading to a Doppler deboosted indirect image on the right. In the bottom right snapshot, one sees the Doppler deboosted direct image of the flux eruption on the right, leading to a Doppler boosted indirect image on the left. The Doppler boosting and deboosting phases (which refer to the indirect image) are also demarcated with arrows in Figure~\ref{fig:semifig}\footnote{The effects of black hole spin on these conclusions are minimal, and these results are very similar for a Schwarzschild ($a=0$) black hole. In the Schwarzschild case, Doppler effects vanish completely, as the zero-angular momentum frame is motionless everywhere.}

We next return to the flaring GRMHD simulations to evaluate whether the gross behavior of the toy model is seen under more physically plausible circumstances.

\section{GRMHD}\label{sec:grmhdsec}
In this section, we describe the GRMHD simulation and ray-tracing techniques that we use to measure subring fluxes in various accreting environments.

\subsection{Procedure}
To further investigate the connection between photon ring brightness and flux eruption events, we ray-trace the ideal GRMHD simulation from \citet{Ripperda_2022}, performed using the \texttt{h-amr} code \citep[][]{Liska_hamr}. The simulation shows MAD accretion cycles, separated by prominent plasmoid-mediated magnetic reconnection events through which magnetic flux is expelled from the event horizon. The dimensionless black hole spin parameter is $a=15/16$ and the effective grid resolution is $ N_r\times N_{\theta} \times N_{\phi} = 5376 \times 2304 \times 2304$ defined for logarithmic Kerr-Schild spherical polar coordinates. The simulation was evolved to $t=10,000M$.

We ray-trace the simulation using the adaptive sampling scheme of \texttt{ipole}. We used the mass and distance of M87$^*$, as for the toy model discussed in Section~\ref{sec:toymodel}, but we reduce the FOV to $80\,\mu$as. The GRMHD scale factor is calibrated so that the average flux density is $F_{230}\sim 0.5$\,Jy. In generating the images, we rotated our azimuthal coordinates clockwise by $150^\circ$ from the \texttt{ipole} default to align the region of highest synchrotron emissivity with the observer at $\phi_{\rm cam}=0$, following the conventions of the toy model described in Section~\ref{sec:toymodel}.

Over the course of the simulation, the flow exhibits a quiescent state (wherein material accretes steadily onto the black hole) and a flaring state (wherein magnetic flux is expelled outward as described in Section~\ref{sec:intro}). We isolate two GRMHD time slices of the simulation that directly showcase these different states: $t=8858 M$ is identified with quiescent accretion, and $t=9553 M$ is identified with the flux eruption event. During the latter state, roughly half the fluid cells have radially inward velocities (hence accreting) while half have radially outward velocities (hence ejecting). The structure of the flux eruption in the GRMHD simulation is thus broadly consistent with our choice of toy model in Section~\ref{sec:toymodel}. 

We note that not only are there many smaller-scale eruption events that occur during the GRMHD simulation, but the size of the ejection region for any one specific eruption event also changes over time. We used our toy model to represent the peak of a particularly large flux eruption event from this simulation so as to capture the near-horizon image structure during a potentially bright high-energy flare.

\subsection{Photon Ring Flux Ratios}
As with the toy model, to a leading order approximation, we can ray-trace a single time-slice of data for a range of values of $\phi_{\rm cam}$ as a proxy for tracking the eruption over time. In particular, this eliminates the need to account for the time-dependence of the eruption shape, which would introduce additional non-linearities into our analysis.  

Unlike the toy model, however, the GRMHD data contains information about magnetic fields. We expect the orientation of these fields in the accretion flow to directly control the brightness of the photon ring; for synchrotron processes, the plasma emissivity depends on $\vec{k}\times \vec{B}$, where $\vec{k}$ is the spatial 3-vector of the null geodesic and $\vec{B}$ is the magnetic field, both of which are measured in the local Minkowski frame of the source \citep{Narayan_2021}. Indeed, for M87$^*$ and Sgr~A$^*$ at millimeter wavelengths, the specific intensity will be roughly proportional to ${\sim}\sin^2\zeta$, where $\zeta$ is the ``pitch angle" between the emitted wavevector and the magnetic field \citep[see, e.g.,][]{Narayan_2021}. 

\begin{figure*}
\centering
    \includegraphics[width=\textwidth]{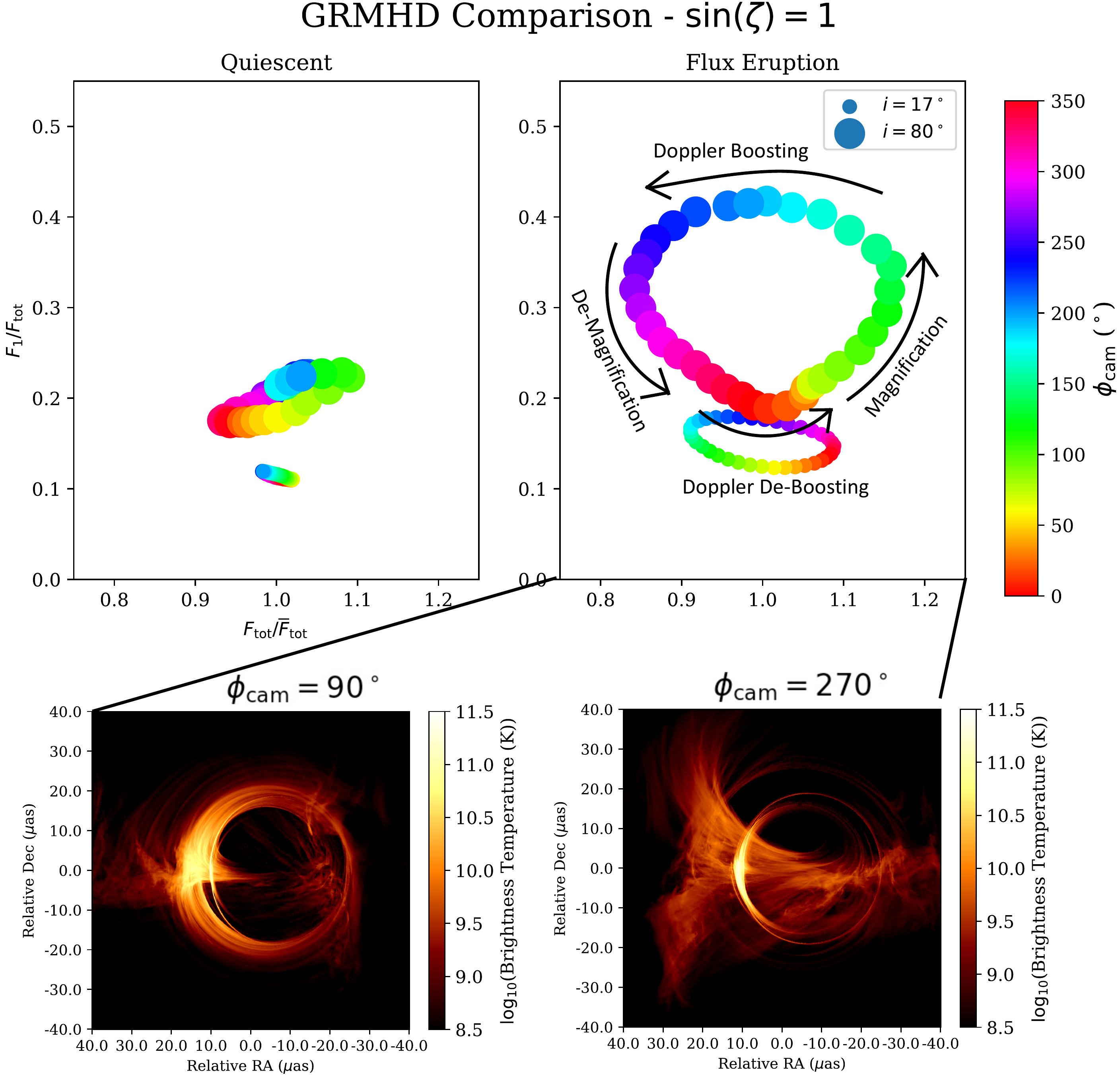}
	\caption{GRMHD PFR's after removing magnetic field directional dependence. Snapshots (bottom) are for $\theta_{\rm cam}=80^\circ$. Here, ``quiescent" and ``flux eruption" refer to time slices $t=8858 M$ and $t=9553 M$ respectively. During the flux eruption, the shape of the high-inclination $(\theta_{\rm cam}=80^\circ)$ PFR curve closely resembles that of the toy model, driven by the effects of Doppler boosting and magnification. Furthermore, the direct and indirect images of the flux eruption can each be seen in the snapshots below.}
    \label{fig:grmhddirection}
\end{figure*}

To compare the GRMHD case to Figure~\ref{fig:semifig}, we first ray-trace the data for a range of $\phi_{\rm cam}$ values
with $\sin\zeta \equiv 1$ everywhere, thus eliminating magnetic field directional dependence and causing the GRMHD emissivity prescription to resemble that of the toy model. We then construct curves of the GRMHD PFR as a function of $\phi_{\rm cam}$ for both $\theta_{\rm cam}=17^\circ$ and $\theta_{\rm cam}=80^\circ$, and the results are plotted in Figure~\ref{fig:grmhddirection}.

Qualitatively, the high-inclination curves of Figure~\ref{fig:grmhddirection} match those of semianalytic case in Figure~\ref{fig:semifig}. All four phases (magnification, boosting, demagnification, and deboosting) are evident. Also evident are the Doppler boosted direct image of the eruption in the lower left panel of Figure~\ref{fig:grmhddirection}, as well as the Doppler boosted indirect image of the eruption in the lower right panel Figure~\ref{fig:grmhddirection}.

However, the low-inclination curve now rotates in the opposite direction (i.e., clockwise corresponds to \emph{increasing} $\phi_{\rm cam}$) and has been stretched out dramatically, indicating a break in axisymmetry; the region emitting in the sub-millimeter has been restricted to a narrower range of $\phi$ values. The effects of magnetic field magnitude are primarily responsible for this change.

Next, we will investigate the role of the magnetic field direction by re-introducing the $\sin\zeta$ dependence in the synchrotron emissivity.

\section{Magnetic Fields}\label{sec:magnetism}
In this section, we describe the specific role that magnetic field direction plays in altering the relative intensity of the direct vs. indirect images.

\subsection{Background}\label{sec:background}
The pitch angle $\zeta$, which encodes the directional dependence of synchrotron emissivity, can dramatically alter the brightness of the image in a way that cannot be predicted from properties of the spacetime alone. In particular, $\zeta$ is \emph{different} for the direct and indirect images and therefore directly influences the PFR. For many field configurations, the pitch angle will be larger for the strongly lensed geodesics and can hence artificially inflate the relative brightness of the photon ring. 

For M87$^*$, it is believed that the accretion flow has a strongly poloidal (i.e., a mix of vertical and radial) magnetic field and that the hole is viewed from Earth at a relatively low inclination \citep{EHT_8,EHT_1}. For instance, for a stationary emitter at the ISCO ($r=6 M$) of a Schwarzschild black hole viewed at face-on inclination in a purely vertical magnetic field, the pitch angle is $\zeta=19.4^\circ$ for the $n=0$ image and $\zeta=41.8^\circ$ for the $n=1$ image. In this case, $\sin^2\zeta$ increases by more than a factor of 4 in the indirect image, leading to a corresponding increase in the relative brightness of the photon ring. 

Such differences in field configurations are particularly relevant to our discussion of flux eruption events. During the quiescent state, the magnetic field is less ordered and accretes together with the infalling gas. However, during the flaring state, the magnetic field in the remaining magnetosphere (after material has been ejected) is connected to the event horizon and the jet, rather than the disk. This causes the magnetic field to be nearly equatorial and contain a magnetic null (i.e., a current sheet where reconnection takes place) separating the northern and southern jets. These contrasting magnetic field configurations may result in observable differences in direct emission that may be more prominent in polarized emission. 

\subsection{Effects of Magnetic Field Direction on GRMHD PFR's}
\begin{figure*}
\centering
    \includegraphics[width=\textwidth]{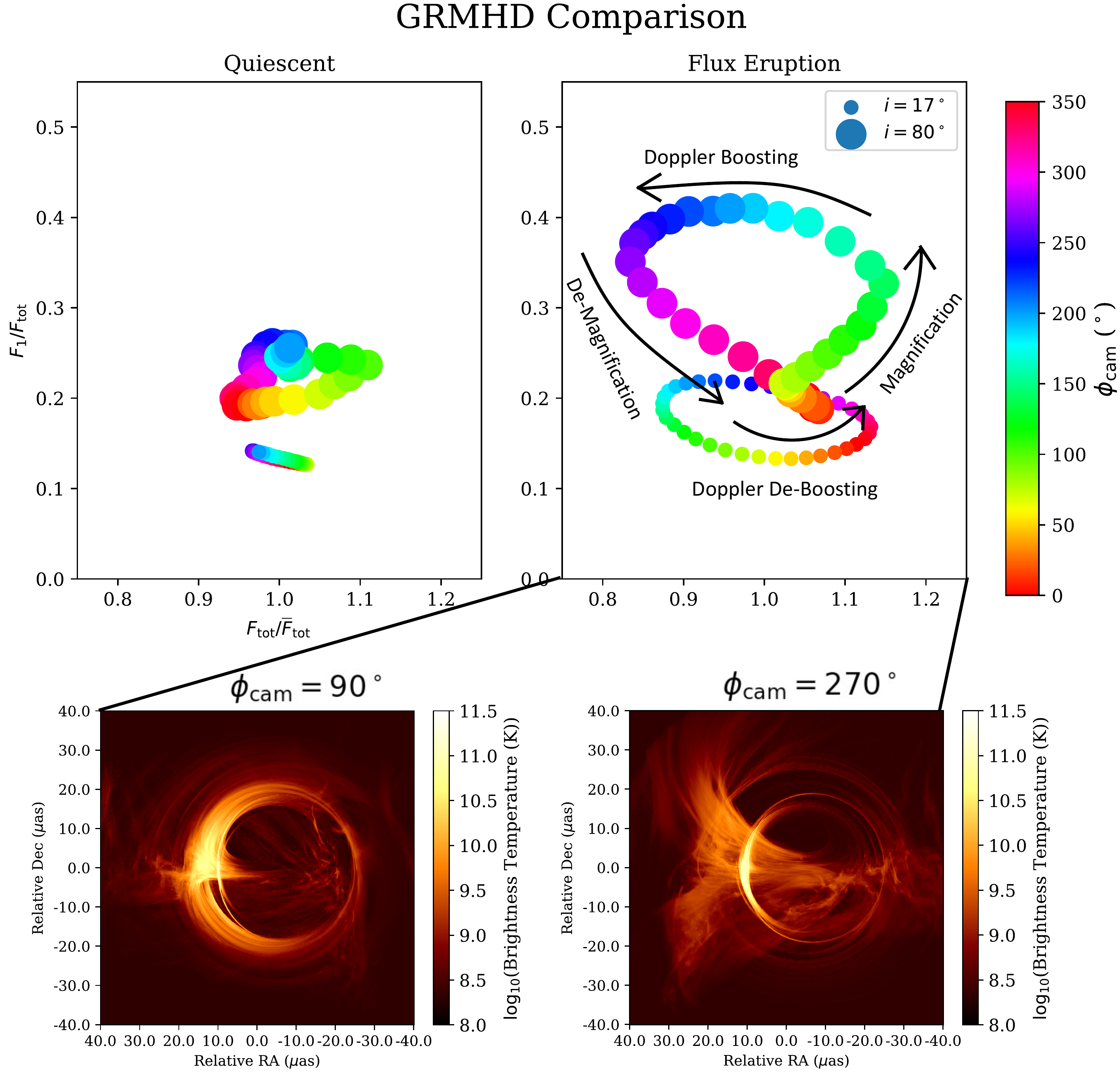}
	\caption{PFR curves for GRMHD ray-traced with full synchrotron emissivities (i.e., including magnetic field dependence). The results are similar to \autoref{fig:grmhddirection}, showing that the effects of magnetic field direction are insignificant in shaping the PFR for this example both in the quiescent state and during the flux eruption.
	}
    \label{fig:comparegrmhd}
\end{figure*}

To demonstrate the effects of magnetic field direction on the observed PFR's, we ray-trace the GRMHD simulation anew with a synchrotron emissivity profile that depends appropriately on $\sin\zeta$ (as would normally be done). The resultant PFR's are shown in Figure~\ref{fig:comparegrmhd}. Qualitatively, Figure~\ref{fig:comparegrmhd} matches Figure~\ref{fig:grmhddirection}. All four phases are once again identifiable in the high-inclination PFR curve, although the Doppler deboosting phase has shrunk to a smaller region of the phase space.

The low-inclination curve, while similar in orientation to Figure~\ref{fig:grmhddirection}, has been dilated and translated upward, indicating the presence of a brighter photon ring. This transformation is consistent with a predominantly vertical magnetic field configuration in the accretion flow. As explained in Section~\ref{sec:background}, the contrast in pitch angle between direct/indirect images is particularly strong when the magnetic field threading the disk points directly at (or directly away from) the observer, as is the case for a vertical field viewed nearly face-on. In this case, we expect the PFR to increase, and we subsequently expect the behavior seen in Figure~\ref{fig:comparegrmhd}.

The correspondence between Figures~\ref{fig:semifig}, \ref{fig:grmhddirection}, and~\ref{fig:comparegrmhd} suggest that the half-disk is an appropriate (albeit simplistic) toy model for broadly representing the shape of a flux eruption event, as well as its signatures of gravitational lensing and Doppler effects. However, a dynamically important magnetic field is necessary to fully capture the correct emissivity profile and resultant PFR.

\section{Discussion}\label{sec:discussion}

We have analyzed sub-millimeter images of black holes from semi-analytic toy models and GRMHD simulations with plasmoid-mediated magnetic reconnection events. We have shown that these flux eruption events illuminate clear signatures of light bending near a black hole, and that these specific signatures depend on a multitude of factors stemming from both the spacetime and astrophysics at hand. In particular, at sub-millimeter wavelengths, the relative brightness of the $n=1$ photon ring, $f_1$, is sensitive to the orbital phase of the flux eruption and can exceed 40\% in systems viewed at high inclination. Furthermore, we have shown that during these flux eruption events, the PFR traces out a generic looping pattern over time, which is due to a combination of gravitational lensing, Doppler boosting, and changing synchrotron emissivity. Each of these factors is significant. \medskip

These striking features in the time-variable images of black holes are an exceptional opportunity for studies with the next generation EHT (ngEHT). In particular, the ngEHT will enable time-resolved images of both Sgr~A$^*$ and M87$^*$ over hundreds-to-thousands of gravitational timescales, sufficient to catch rare events and to monitor their evolution. The enhanced angular resolution, baseline coverage, and sensitivity of the ngEHT will enable detailed studies of the photon ring \citep[see, e.g.,][]{Tiede_2022}, allowing measurements of the PFR during flux eruptions and flares\footnote{By flares, we refer to transient bright emission usually observed at wavelengths much smaller than sub-millimeter, e.g., X-ray flares in Sgr A$^*$ \citep[][]{Haggard:2019}.}. Finally, simultaneous multi-frequency capabilities of the ngEHT will allow for spectral index and polarization measurements during flares \citep[see, e.g.,][]{Ricarte_2022}, providing additional information that can resolve the degeneracies between effects from the curved spacetime and emitting plasma. Along with the ngEHT, simultaneous multiwavelength coverage at higher energies such as near-infrared and X-rays for Sgr A$^*$ \citep[e.g.,][]{GRAVITY:2021,Boyce:2022} or TeV for M87$^*$ \citep[e.g.,][]{Aharonian:2006} could constrain both the flow structure and particle acceleration mechanisms during eruption events.\medskip 

We emphasize that our model does not capture several intricacies of the problem at hand. While our moving camera does serve as an effective time-proxy for the rotation of the accretion flow, we employ a ``fast-light" ray tracing algorithm that does not take into account the time delay between direct and indirect images. While ``slow-light" simulations are significantly more computationally expensive, time delays can be integral to our study of hotspot lightcurves and images \citep[e.g.,][]{Broderick_2005,Broderick_2006}, so they should be employed in future studies of the PFR. Qualitatively, we suspect that the effects of time delay on the PFR curves are two-fold. First, the curve should rotate, as the point of maximal/minimal lensing will be shifted. Second, the curve should shrink, as direct/indirect images will no longer be diametrically opposed on the observer's screen, leading to a smaller range of possible values for $f_1$. In any case, we do not expect the time delay to modify our conclusions, as the generic shape of the PFR curve will remain consistent.\medskip 

This study has provided a first glimpse of the scientific opportunities that may be possible with time-resolved studies of the photon ring during flux eruptions of magnetically arrested disks. Additional crucial topics for future studies include the accessible signatures in polarization, the effects of optical depth, and reconstructed movies with the ngEHT. 

\vspace{6pt} 

\funding{We thank the National Science Foundation (AST-1716536 and AST-1935980) and the Gordon and Betty Moore Foundation (GBMF-10423) for financial support of this work. This work was supported in part by the Black Hole Initiative, which is funded by grants from the John Templeton Foundation and the Gordon and Betty Moore Foundation to Harvard University. K.C. is also supported in part by the Black Hole PIRE program (NSF grant OISE-1743747). B.R. is supported by a Joint Princeton/Flatiron Postdoctoral Fellowship. Research at the Flatiron Institute is supported by the Simons Foundation. M.L. is supported by John Harvard Distinguished Science Fellowship and ITC Fellowship.}

\dataavailability{Simulation and raytraced data available on request to authors. Software: eht-imaging library \citep{Chael_2016}, \texttt{h-amr} \citep[][]{Liska_hamr}, \texttt{ipole} \citep[][]{ipole_2018}, Numpy \citep{numpy}, Matplotlib \citep{matplotlib}} 

\authorcontributions{Conceptualization, Z.G., K.C. and M.J.; methodology, Z.G. and K.C.; software, Z.G., K.C. and M.L.; formal analysis, Z.G. and K.C.; writing---original draft preparation, Z.G., K.C. and M.J.; writing---review and editing, Z.G., K.C., M.J., B.R. and M.L.; visualization, Z.G., K.C. and M.J.; supervision, K.C. and M.J. All authors have read and agreed to the published version of the manuscript.}

\acknowledgments{We thank the referees for their comments and suggestions. We thank Maciek Wielgus and Dominic Chang for useful discussions. This research was enabled by support provided by grant no. NSF PHY-1125915 along with a INCITE program award PHY129, using resources from the Oak Ridge Leadership Computing Facility, Summit, which is a US Department of Energy office of Science User Facility supported under contract DE-AC05- 00OR22725, as well as Calcul Quebec (http://www.calculquebec.ca) and Compute Canada (http://www.computecanada.ca). The computational resources and services used in this work were partially provided by facilities supported by the Scientific Computing Core at the Flatiron Institute, a division of the Simons Foundation. This research is part of the Frontera (\citealt{Frontera}) computing project at the Texas Advanced Computing Center (LRAC-AST20008). Frontera is made possible by National Science Foundation award OAC-1818253.}

\conflictsofinterest{The authors declare no conflict of interest.}

\abbreviations{Abbreviations}{
The following abbreviations are used in this manuscript:\\

\noindent 
\begin{tabular}{@{}ll}
BH & Black Hole\\
EHT & Event Horizon Telescope\\
FOV & Field Of View\\
GRMHD & General Relativistic Magneto Hydro Dynamic \\
GRRT & General Relativistic Ray Tracing\\
ISCO & Inner Stable Circular Orbit\\
MAD & Magnetically Arrested Disk\\
ngEHT & next generation Event Horizon Telescope\\
PFR & Photon ring Flux Ratio\\
SMBH & Super Massive Black Hole\\
VLBI & Very Long Baseline Interferometry

\end{tabular}
}

%%%%%%%%%%%%%%%%%%%%%%%%%%%%%%%%%%%%%%%%%%
%% Optional
\appendixtitles{yes} % Leave argument "no" if all appendix headings stay EMPTY (then no dot is printed after "Appendix A"). If the appendix sections contain a heading then change the argument to "yes".
\appendixstart
\appendix
\section{Magnetic flux eruptions in GRMHD}\label{sec:grmhdapp}

\begin{figure*}
    \includegraphics[width=\textwidth]{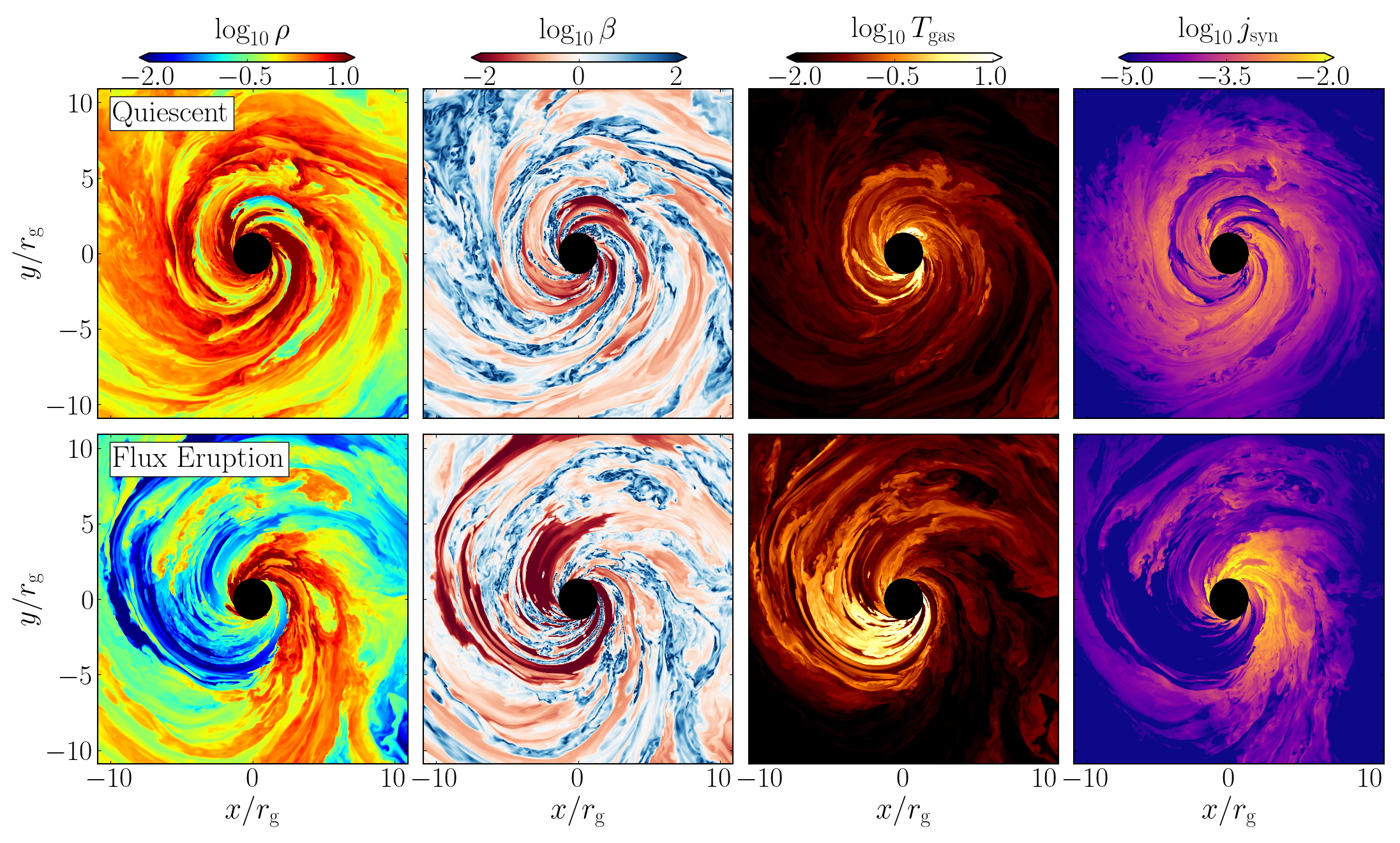}
	\caption{Magnetic flux eruptions produce regions of relativistically hot gas that could potentially produce high-energy flares in Sgr A$^*$ and M87$^*$. This figure shows midplane cross-sections of the quiescent (top) and flux eruption (bottom) states from the GRMHD simulation. From left-to-right, panels show the gas density $\rho$, plasma-$\beta$, gas temperature $T_{\rm gas}$ (in relativistic units), and a proxy for the 230\,GHz synchrotron emissivity $j_{\rm syn}$. 
 }
    \label{fig:GRMHD}
\end{figure*}

From Fig.~\ref{fig:comparefig}, we see that the gas inspirals towards the black hole during the quiescent state ($t=8858M$), while roughly half of the disk is ejected during the flux eruption event ($t=9553M$). Here we briefly discuss the changes in other properties of the accretion flow brought about by magnetic flux eruptions. Figure~\ref{fig:GRMHD} shows the midplane cross-section of the gas density $\rho$, plasma-$\beta$ (i.e., the ratio of the thermal and magnetic pressures, $p_{\rm gas}/p_{\rm mag}$), and gas temperature $T_{\rm gas}=p_{\rm gas}/\rho$. We also show the proxy for the 230 GHz thermal synchrotron emissivity $j_{\rm syn}$ given by the EHT code comparison project \citet{Porth_2019}:
\begin{equation}
    j_{\rm syn}=\frac{\rho^3}{p_{\rm gas}^2}\exp \left[-C \left(\frac{\rho^2}{B p_{\rm gas}^2}\right)^{1/3}\right].
\end{equation}
This emissivity prescription is designed to resemble the true synchrotron emissivity fitting function given in \citet{Leung:2011}. Following \citet{Porth_2019}, we assume $C=0.2$ such that the emission drops exponentially beyond a few gravitational radii. We further normalize $j_{\rm syn}$ such that the volume-integrated total synchrotron emissivity is 1.

The quiescent state is characterized by gas-rich spiral features that interact with strong magnetic fields (indicated by $\beta<1$) near the black hole. There are occasional reconnecting sites where magnetic dissipation leads to relativistic gas temperatures $(i.e., T_{\rm gas}>1)$. overall, we see a roughly axisymmetric disk structure that produces an azimuthally-uniform emissivity profile.

On the other hand, the flux eruption state exhibits highly relativistic temperatures in the evacuated region. This occurs due to the formation of a long thin current sheet that ultimately destabilizes and undergoes reconnection, allowing the escape of a magnetic flux-tube \citep[][]{Ripperda_2022}. The large gas temperatures could potentially produce high-energy emission that may be able to explain X-ray/$\gamma-$ray flares seen in low-luminosity super-massive black holes such as Sgr A$^*$ and M87$^*$. For the 230 GHz image, the synchrotron emissivity map predicts that the bulk of the emission originates in the highly-dense relatively low-temperature accreting region. The evacuated region, despite the high gas temperatures, produces little to no flux, leading to the unique azimuthally-dependent image morphologies shown in Fig.~\ref{fig:comparegrmhd}.

\begin{adjustwidth}{-\extralength}{0cm}
%\printendnotes[custom] % Un-comment to print a list of endnotes

\reftitle{References}

\bibliography{bib}

\begin{thebibliography}{999}

\bibitem[{Event Horizon Telescope Collaboration} \em{et~al.}({2019a}){Event
  Horizon Telescope Collaboration}, {Akiyama}, {Alberdi}, {Alef}, {Asada},
  {Azulay}, {Baczko}, {Ball}, {Balokovi{\'c}}, {Barrett}, and et~al.]{EHT_1}
{Event Horizon Telescope Collaboration}.; {Akiyama}, K.; {Alberdi}, A.; {Alef},
  W.; {Asada}, K.; {Azulay}, R.; {Baczko}, A.K.; {Ball}, D.; {Balokovi{\'c}},
  M.; {Barrett}, J.;  et~al.
\newblock {First M87 Event Horizon Telescope Results. I. The Shadow of the
  Supermassive Black Hole}.
\newblock {\em \apjl} {\bf {2019a}}, {\em 875},~L1,
  \href{http://xxx.lanl.gov/abs/1906.11238}{{\normalfont [1906.11238]}}.
\newblock {\url{https://doi.org/10.3847/2041-8213/ab0ec7}}.

\bibitem[{Event Horizon Telescope Collaboration} \em{et~al.}(2022){Event
  Horizon Telescope Collaboration}, {Akiyama}, {Alberdi}, {Alef}, {Algaba},
  {Anantua}, {Asada}, {Azulay}, {Bach}, {Baczko}, {Ball}, {Balokovi{\'c}},
  {Barrett}, {Baub{\"o}ck}, {Benson}, {Bintley}, {Blackburn}, {Blundell},
  {Bouman}, {Bower}, {Boyce}, {Bremer}, {Brinkerink}, {Brissenden}, {Britzen},
  {Broderick}, {Broguiere}, {Bronzwaer}, {Bustamante}, {Byun}, {Carlstrom},
  {Ceccobello}, {Chael}, {Chan}, {Chatterjee}, {Chatterjee}, {Chen}, {Chen},
  {Cheng}, {Cho}, {Christian}, {Conroy}, {Conway}, {Cordes}, {Crawford},
  {Crew}, {Cruz-Osorio}, {Cui}, {Davelaar}, {Laurentis}, {Deane}, {Dempsey},
  {Desvignes}, {Dexter}, {Dhruv}, {Doeleman}, {Dougal}, {Dzib}, {Eatough},
  {Emami}, {Falcke}, {Farah}, {Fish}, {Fomalont}, {Ford}, {Fraga-Encinas},
  {Freeman}, {Friberg}, {Fromm}, {Fuentes}, {Galison}, {Gammie}, {Garc{\'\i}a},
  {Gentaz}, {Georgiev}, {Goddi}, {Gold}, {G{\'o}mez-Ruiz}, {G{\'o}mez}, {Gu},
  {Gurwell}, {Hada}, {Haggard}, {Haworth}, {Hecht}, {Hesper}, {Heumann}, {Ho},
  {Ho}, {Honma}, {Huang}, {Huang}, {Hughes}, {Ikeda}, {Impellizzeri}, {Inoue},
  {Issaoun}, {James}, {Jannuzi}, {Janssen}, {Jeter}, {Jiang},
  {Jim{\'e}nez-Rosales}, {Johnson}, {Jorstad}, {Joshi}, {Jung}, {Karami},
  {Karuppusamy}, {Kawashima}, {Keating}, {Kettenis}, {Kim}, {Kim}, {Kim},
  {Kim}, {Kino}, {Koay}, {Kocherlakota}, {Kofuji}, {Koch}, {Koyama}, {Kramer},
  {Kramer}, {Krichbaum}, {Kuo}, {Bella}, {Lauer}, {Lee}, {Lee}, {Leung},
  {Levis}, {Li}, {Lico}, {Lindahl}, {Lindqvist}, {Lisakov}, {Liu}, {Liu},
  {Liuzzo}, {Lo}, {Lobanov}, {Loinard}, {Lonsdale}, {Lu}, {Mao}, {Marchili},
  {Markoff}, {Marrone}, {Marscher}, {Mart{\'\i}-Vidal}, {Matsushita},
  {Matthews}, {Medeiros}, {Menten}, {Michalik}, {Mizuno}, {Mizuno}, {Moran},
  {Moriyama}, {Moscibrodzka}, {M{\"u}ller}, {Mus}, {Musoke}, {Myserlis},
  {Nadolski}, {Nagai}, {Nagar}, {Nakamura}, {Narayan}, {Narayanan},
  {Natarajan}, {Nathanail}, {Fuentes}, {Neilsen}, {Neri}, {Ni}, {Noutsos},
  {Nowak}, {Oh}, {Okino}, {Olivares}, {Ortiz-Le{\'o}n}, {Oyama}, {{\"O}zel},
  {Palumbo}, {Paraschos}, {Park}, {Parsons}, {Patel}, {Pen}, {Pesce},
  {Pi{\'e}tu}, {Plambeck}, {PopStefanija}, {Porth}, {P{\"o}tzl}, {Prather},
  {Preciado-L{\'o}pez}, {Psaltis}, {Pu}, {Ramakrishnan}, {Rao}, {Rawlings},
  {Raymond}, {Rezzolla}, {Ricarte}, {Ripperda}, {Roelofs}, {Rogers}, {Ros},
  {Romero-Ca{\~n}izales}, {Roshanineshat}, {Rottmann}, {Roy}, {Ruiz},
  {Ruszczyk}, {Rygl}, {S{\'a}nchez}, {S{\'a}nchez-Arg{\"u}elles},
  {S{\'a}nchez-Portal}, {Sasada}, {Satapathy}, {Savolainen}, {Schloerb},
  {Schonfeld}, {Schuster}, {Shao}, {Shen}, {Small}, {Sohn}, {SooHoo},
  {Souccar}, {Sun}, {Tazaki}, {Tetarenko}, {Tiede}, {Tilanus}, {Titus},
  {Torne}, {Traianou}, {Trent}, {Trippe}, {Turk}, {van Bemmel}, {van
  Langevelde}, {van Rossum}, {Vos}, {Wagner}, {Ward-Thompson}, {Wardle},
  {Weintroub}, {Wex}, {Wharton}, {Wielgus}, {Wiik}, {Witzel}, {Wondrak},
  {Wong}, {Wu}, {Yamaguchi}, {Yoon}, {Young}, {Young}, {Younsi}, {Yuan},
  {Yuan}, {Zensus}, {Zhang}, {Zhao}, {Zhao}, {Agurto}, {Allardi}, {Amestica},
  {Araneda}, {Arriagada}, {Berghuis}, {Bertarini}, {Berthold}, {Blanchard},
  {Brown}, {C{\'a}rdenas}, {Cantzler}, {Caro}, {Castillo-Dom{\'\i}nguez},
  {Chan}, {Chang}, {Chang}, {Chang}, {Chang}, {Chen}, {Chilson}, {Chuter},
  {Ciechanowicz}, {Colin-Beltran}, {Coulson}, {Crowley}, {Degenaar},
  {Dornbusch}, {Dur{\'a}n}, {Everett}, {Faber}, {Forster}, {Fuchs}, {Gale},
  {Geertsema}, {Gonz{\'a}lez}, {Graham}, {Gueth}, {Halverson}, {Han}, {Han},
  {Hasegawa}, {Hern{\'a}ndez-Rebollar}, {Herrera}, {Herrero-Illana},
  {Heyminck}, {Hirota}, {Hoge}, {Hostler Schimpf}, {Howie}, {Huang}, {Jiang},
  {Jinchi}, {John}, {Kimura}, {Klein}, {Kubo}, {Kuroda}, {Kwon}, {Lacasse},
  {Laing}, {Leitch}, {Li}, {Liu}, {Liu}, {Lin}, {Lu}, {Mac-Auliffe},
  {Martin-Cocher}, {Matulonis}, {Maute}, {Messias}, {Meyer-Zhao},
  {Monta{\~n}a}, {Montenegro-Montes}, {Montgomerie}, {Moreno Nolasco},
  {Muders}, {Nishioka}, {Norton}, {Nystrom}, {Ogawa}, {Olivares}, {Oshiro},
  {P{\'e}rez-Beaupuits}, {Parra}, {Phillips}, {Poirier}, {Pradel}, {Qiu},
  {Raffin}, {Rahlin}, {Ram{\'\i}rez}, {Ressler}, {Reynolds},
  {Rodr{\'\i}guez-Montoya}, {Saez-Madain}, {Santana}, {Shaw}, {Shirkey},
  {Silva}, {Snow}, {Sousa}, {Sridharan}, {Stahm}, {Stark}, {Test},
  {Torstensson}, {Venegas}, {Walther}, {Wei}, {White}, {Wieching}, {Wijnands},
  {Wouterloot}, {Yu}, {Yu (于威)}, and {Zeballos}]{EHT_sgr_1}
{Event Horizon Telescope Collaboration}.; {Akiyama}, K.; {Alberdi}, A.; {Alef},
  W.; {Algaba}, J.C.; {Anantua}, R.; {Asada}, K.; {Azulay}, R.; {Bach}, U.;
  {Baczko}, A.K.;  et~al.
\newblock {First Sagittarius A* Event Horizon Telescope Results. I. The Shadow
  of the Supermassive Black Hole in the Center of the Milky Way}.
\newblock {\em \apjl} {\bf 2022}, {\em 930},~L12.
\newblock
  {\url{https://doi.org/10.3847/2041-8213/ac667410.3847/2041-8213/ac667510.3847/2041-8213/ac6429}}.

\bibitem[{Luminet}(1979)]{Luminet_1979}
{Luminet}, J.P.
\newblock {Image of a spherical black hole with thin accretion disk}.
\newblock {\em \aap} {\bf 1979}, {\em 75},~228--235.

\bibitem[{de Vries}(2000)]{deVries_2000}
{de Vries}, A.
\newblock {The apparent shape of a rotating charged black hole, closed photon
  orbits and the bifurcation set A$_{4}$}.
\newblock {\em Classical and Quantum Gravity} {\bf 2000}, {\em 17},~123--144.
\newblock {\url{https://doi.org/10.1088/0264-9381/17/1/309}}.

\bibitem[{Takahashi}(2004)]{Takahashi_2004}
{Takahashi}, R.
\newblock {Shapes and Positions of Black Hole Shadows in Accretion Disks and
  Spin Parameters of Black Holes}.
\newblock {\em \apj} {\bf 2004}, {\em 611},~996--1004,
  \href{http://xxx.lanl.gov/abs/astro-ph/0405099}{{\normalfont
  [arXiv:astro-ph/astro-ph/0405099]}}.
\newblock {\url{https://doi.org/10.1086/422403}}.

\bibitem[{Beckwith} and {Done}(2005)]{Beckwith_2005}
{Beckwith}, K.; {Done}, C.
\newblock {Extreme gravitational lensing near rotating black holes}.
\newblock {\em \mnras} {\bf 2005}, {\em 359},~1217--1228,
  \href{http://xxx.lanl.gov/abs/astro-ph/0411339}{{\normalfont
  [arXiv:astro-ph/astro-ph/0411339]}}.
\newblock {\url{https://doi.org/10.1111/j.1365-2966.2005.08980.x}}.

\bibitem[{Johannsen} and {Psaltis}(2010)]{Johannsen_Psaltis_2010}
{Johannsen}, T.; {Psaltis}, D.
\newblock {Testing the No-hair Theorem with Observations in the Electromagnetic
  Spectrum. II. Black Hole Images}.
\newblock {\em \apj} {\bf 2010}, {\em 718},~446--454,
  \href{http://xxx.lanl.gov/abs/1005.1931}{{\normalfont
  [arXiv:astro-ph.HE/1005.1931]}}.
\newblock {\url{https://doi.org/10.1088/0004-637X/718/1/446}}.

\bibitem[{Gralla} \em{et~al.}(2019){Gralla}, {Holz}, and {Wald}]{Gralla_2019}
{Gralla}, S.E.; {Holz}, D.E.; {Wald}, R.M.
\newblock {Black hole shadows, photon rings, and lensing rings}.
\newblock {\em \prd} {\bf 2019}, {\em 100},~024018,
  \href{http://xxx.lanl.gov/abs/1906.00873}{{\normalfont
  [arXiv:astro-ph.HE/1906.00873]}}.
\newblock {\url{https://doi.org/10.1103/PhysRevD.100.024018}}.

\bibitem[Johnson \em{et~al.}(2020)Johnson, Lupsasca, Strominger, Wong, Hadar,
  Kapec, Narayan, Chael, Gammie, Galison, and et~al.]{Johnson_2020}
Johnson, M.D.; Lupsasca, A.; Strominger, A.; Wong, G.N.; Hadar, S.; Kapec, D.;
  Narayan, R.; Chael, A.; Gammie, C.F.; Galison, P.;  et~al.
\newblock Universal interferometric signatures of a black hole’s photon ring.
\newblock {\em Science Advances} {\bf 2020}, {\em 6},~eaaz1310.
\newblock {\url{https://doi.org/10.1126/sciadv.aaz1310}}.

\bibitem[{Darwin}(1959)]{Darwin_1959}
{Darwin}, C.
\newblock {The Gravity Field of a Particle}.
\newblock {\em Proceedings of the Royal Society of London Series A} {\bf 1959},
  {\em 249},~180--194.
\newblock {\url{https://doi.org/10.1098/rspa.1959.0015}}.

\bibitem[{Ohanian}(1987)]{Ohanian_1987}
{Ohanian}, H.C.
\newblock {The black hole as a gravitational ``lens''}.
\newblock {\em American Journal of Physics} {\bf 1987}, {\em 55},~428--432.
\newblock {\url{https://doi.org/10.1119/1.15126}}.

\bibitem[{Broderick} \em{et~al.}(2020){Broderick}, {Pesce}, {Tiede}, {Pu}, and
  {Gold}]{Broderick_2020}
{Broderick}, A.E.; {Pesce}, D.W.; {Tiede}, P.; {Pu}, H.Y.; {Gold}, R.
\newblock {Hybrid Very Long Baseline Interferometry Imaging and Modeling with
  THEMIS}.
\newblock {\em \apj} {\bf 2020}, {\em 898},~9.
\newblock {\url{https://doi.org/10.3847/1538-4357/ab9c1f}}.

\bibitem[{Tiede} \em{et~al.}(2022){Tiede}, {Broderick}, {Palumbo}, and
  {Chael}]{Tiede_2022}
{Tiede}, P.; {Broderick}, A.E.; {Palumbo}, D.C.M.; {Chael}, A.
\newblock {Measuring the Ellipticity of M 87* Images}.
\newblock {\em arXiv e-prints} {\bf 2022}, p. arXiv:2210.13499,
  \href{http://xxx.lanl.gov/abs/2210.13499}{{\normalfont
  [arXiv:astro-ph.HE/2210.13499]}}.

\bibitem[{Porth} \em{et~al.}(2019){Porth}, {Chatterjee}, {Narayan}, {Gammie},
  {Mizuno}, {Anninos}, {Baker}, {Bugli}, {Chan}, {Davelaar}, {Del Zanna},
  {Etienne}, {Fragile}, {Kelly}, {Liska}, {Markoff}, {McKinney}, {Mishra},
  {Noble}, {Olivares}, {Prather}, {Rezzolla}, {Ryan}, {Stone}, {Tomei},
  {White}, {Younsi}, {Akiyama}, {Alberdi}, {Alef}, {Asada}, {Azulay}, {Baczko},
  {Ball}, {Balokovi{\'c}}, {Barrett}, {Bintley}, {Blackburn}, {Boland},
  {Bouman}, {Bower}, {Bremer}, {Brinkerink}, {Brissenden}, {Britzen},
  {Broderick}, {Broguiere}, {Bronzwaer}, {Byun}, {Carlstrom}, {Chael},
  {Chatterjee}, {Chen}, {Chen}, {Cho}, {Christian}, {Conway}, {Cordes},
  {Geoffrey}, {Crew}, {Cui}, {De Laurentis}, {Deane}, {Dempsey}, {Desvignes},
  {Doeleman}, {Eatough}, {Falcke}, {Fish}, {Fomalont}, {Fraga-Encinas},
  {Freeman}, {Friberg}, {Fromm}, {G{\'o}mez}, {Galison}, {Garc{\'\i}a},
  {Gentaz}, {Georgiev}, {Goddi}, {Gold}, {Gu}, {Gurwell}, {Hada}, {Hecht},
  {Hesper}, {Ho}, {Ho}, {Honma}, {Huang}, {Huang}, {Hughes}, {Ikeda}, {Inoue},
  {Issaoun}, {James}, {Jannuzi}, {Janssen}, {Jeter}, {Jiang}, {Johnson},
  {Jorstad}, {Jung}, {Karami}, {Karuppusamy}, {Kawashima}, {Keating},
  {Kettenis}, {Kim}, {Kim}, {Kim}, {Kino}, {Koay}, {Patrick}, {Koch}, {Koyama},
  {Kramer}, {Kramer}, {Krichbaum}, {Kuo}, {Lauer}, {Lee}, {Li}, {Li},
  {Lindqvist}, {Liu}, {Liuzzo}, {Lo}, {Lobanov}, {Loinard}, {Lonsdale}, {Lu},
  {MacDonald}, {Mao}, {Marrone}, {Marscher}, {Mart{\'\i}-Vidal}, {Matsushita},
  {Matthews}, {Medeiros}, {Menten}, {Mizuno}, {Moran}, {Moriyama},
  {Moscibrodzka}, {M{\"u}ller}, {Nagai}, {Nagar}, {Nakamura}, {Narayanan},
  {Natarajan}, {Neri}, {Ni}, {Noutsos}, {Okino}, {Oyama}, {{\"O}zel},
  {Palumbo}, {Patel}, {Pen}, {Pesce}, {Pi{\'e}tu}, {Plambeck}, {PopStefanija},
  {Preciado-L{\'o}pez}, {Psaltis}, {Pu}, {Ramakrishnan}, {Rao}, {Rawlings},
  {Raymond}, {Ripperda}, {Roelofs}, {Rogers}, {Ros}, {Rose}, {Roshanineshat},
  {Rottmann}, {Roy}, {Ruszczyk}, {Rygl}, {S{\'a}nchez},
  {S{\'a}nchez-Arguelles}, {Sasada}, {Savolainen}, {Schloerb}, {Schuster},
  {Shao}, {Shen}, {Small}, {Sohn}, {SooHoo}, {Tazaki}, {Tiede}, {Tilanus},
  {Titus}, {Toma}, {Torne}, {Trent}, {Trippe}, {Tsuda}, {van Bemmel}, {van
  Langevelde}, {van Rossum}, {Wagner}, {Wardle}, {Weintroub}, {Wex}, {Wharton},
  {Wielgus}, {Wong}, {Wu}, {Young}, {Young}, {Yuan}, {Yuan}, {Zensus}, {Zhao},
  {Zhao}, {Zhu}, and {Event Horizon Telescope Collaboration}]{Porth_2019}
{Porth}, O.; {Chatterjee}, K.; {Narayan}, R.; {Gammie}, C.F.; {Mizuno}, Y.;
  {Anninos}, P.; {Baker}, J.G.; {Bugli}, M.; {Chan}, C.k.; {Davelaar}, J.;
  et~al.
\newblock {The Event Horizon General Relativistic Magnetohydrodynamic Code
  Comparison Project}.
\newblock {\em \apjs} {\bf 2019}, {\em 243},~26,
  \href{http://xxx.lanl.gov/abs/1904.04923}{{\normalfont
  [arXiv:astro-ph.HE/1904.04923]}}.
\newblock {\url{https://doi.org/10.3847/1538-4365/ab29fd}}.

\bibitem[{Gold} \em{et~al.}(2020){Gold}, {Broderick}, {Younsi}, {Fromm},
  {Gammie}, {Mo{\'s}cibrodzka}, {Pu}, {Bronzwaer}, {Davelaar}, {Dexter},
  {Ball}, {Chan}, {Kawashima}, {Mizuno}, {Ripperda}, {Akiyama}, {Alberdi},
  {Alef}, {Asada}, {Azulay}, {Baczko}, {Balokovi{\'c}}, {Barrett}, {Bintley},
  {Blackburn}, {Boland}, {Bouman}, {Bower}, {Bremer}, {Brinkerink},
  {Brissenden}, {Britzen}, {Broguiere}, {Byun}, {Carlstrom}, {Chael},
  {Chatterjee}, {Chatterjee}, {Chen}, {Chen}, {Cho}, {Christian}, {Conway},
  {Cordes}, {Crew}, {Cui}, {De Laurentis}, {Deane}, {Dempsey}, {Desvignes},
  {Doeleman}, {Eatough}, {Falcke}, {Fish}, {Fomalont}, {Fraga-Encinas},
  {Freeman}, {Friberg}, {G{\'o}mez}, {Galison}, {Garc{\'\i}a}, {Gentaz},
  {Georgiev}, {Goddi}, {Gu}, {Gurwell}, {Hada}, {Hecht}, {Hesper}, {Ho}, {Ho},
  {Honma}, {Huang}, {Huang}, {Hughes}, {Inoue}, {Issaoun}, {James}, {Jannuzi},
  {Janssen}, {Jeter}, {Jiang}, {Jimenez-Rosales}, {Johnson}, {Jorstad}, {Jung},
  {Karami}, {Karuppusamy}, {Keating}, {Kettenis}, {Kim}, {Kim}, {Kim}, {Kino},
  {Koay}, {Koch}, {Koyama}, {Kramer}, {Kramer}, {Krichbaum}, {Kuo}, {Lauer},
  {Lee}, {Li}, {Li}, {Lico}, {Lindqvist}, {Liu}, {Liuzzo}, {Lo}, {Lobanov},
  {Loinard}, {Lonsdale}, {Lu}, {MacDonald}, {Markoff}, {Mao}, {Marrone},
  {Marscher}, {Mart{\'\i}-Vidal}, {Matsushita}, {Matthews}, {Medeiros},
  {Menten}, {Mizuno}, {Moran}, {Moriyama}, {M{\"u}ller}, {Nagai}, {Nakamura},
  {Nagar}, {Narayan}, {Narayanan}, {Natarajan}, {Neri}, {Ni}, {Noutsos},
  {Okino}, {Ortiz-Le{\'o}n}, {Oyama}, {{\"O}zel}, {Palumbo}, {Park}, {Patel},
  {Pen}, {Pesce}, {Plambeck}, {Pi{\'e}tu}, {PopStefanija}, {Porth},
  {Preciado-L{\'o}pez}, {Psaltis}, {Ramakrishnan}, {Rao}, {Rawlings},
  {Raymond}, {Rezzolla}, {Roelofs}, {Rogers}, {Ros}, {Rose}, {Roshanineshat},
  {Rottmann}, {Roy}, {Ruszczyk}, {Rygl}, {S{\'a}nchez},
  {S{\'a}nchez-Arguelles}, {Sasada}, {Savolainen}, {Schuster}, {Schloerb},
  {Shao}, {Shen}, {Small}, {Sohn}, {SooHoo}, {Tiede}, {Tazaki}, {Tilanus},
  {Titus}, {Toma}, {Torne}, {Trent}, {Traianou}, {Trippe}, {Tsuda}, {van
  Langevelde}, {van Bemmel}, {van Rossum}, {Wagner}, {Wardle}, {Wex},
  {Weintroub}, {Wharton}, {Wielgus}, {Wong}, {Wu}, {Yoon}, {Young}, {Young},
  {Yuan}, {Yuan}, {Zensus}, {Zhao}, {Zhao}, {Zhu}, and {Event Horizon Telescope
  Collaboration}]{Gold_2020}
{Gold}, R.; {Broderick}, A.E.; {Younsi}, Z.; {Fromm}, C.M.; {Gammie}, C.F.;
  {Mo{\'s}cibrodzka}, M.; {Pu}, H.Y.; {Bronzwaer}, T.; {Davelaar}, J.;
  {Dexter}, J.;  et~al.
\newblock {Verification of Radiative Transfer Schemes for the EHT}.
\newblock {\em \apj} {\bf 2020}, {\em 897},~148.
\newblock {\url{https://doi.org/10.3847/1538-4357/ab96c6}}.

\bibitem[{Event Horizon Telescope Collaboration} \em{et~al.}({2019e}){Event
  Horizon Telescope Collaboration}, {Akiyama}, {Alberdi}, {Alef}, {Asada},
  {Azulay}, {Baczko}, {Ball}, {Balokovi{\'c}}, {Barrett}, and et~al.]{EHT_5}
{Event Horizon Telescope Collaboration}.; {Akiyama}, K.; {Alberdi}, A.; {Alef},
  W.; {Asada}, K.; {Azulay}, R.; {Baczko}, A.K.; {Ball}, D.; {Balokovi{\'c}},
  M.; {Barrett}, J.;  et~al.
\newblock {First M87 Event Horizon Telescope Results. V. Physical Origin of the
  Asymmetric Ring}.
\newblock {\em \apjl} {\bf {2019e}}, {\em 875},~L5,
  \href{http://xxx.lanl.gov/abs/1906.11242}{{\normalfont [1906.11242]}}.
\newblock {\url{https://doi.org/10.3847/2041-8213/ab0f43}}.

\bibitem[{Event Horizon Telescope Collaboration} \em{et~al.}(2022){Event
  Horizon Telescope Collaboration}, {Akiyama}, {Alberdi}, {Alef}, {Carlos
  Algaba}, {Anantua}, {Asada}, {Azulay}, {Bach}, {Baczko}, {Ball},
  {Balokovi{\'c}}, {Barrett}, {Baub{\"o}ck}, {Benson}, {Bintley}, {Blackburn},
  {Blundell}, {Bouman}, {Bower}, {Boyce}, {Bremer}, {Brinkerink}, {Brissenden},
  {Britzen}, {Broderick}, {Broguiere}, {Bronzwaer}, {Bustamante}, {Byun},
  {Carlstrom}, {Ceccobello}, {Chael}, {Chan}, {Chatterjee}, {Chatterjee},
  {Chen}, {Chen}, {Cheng}, {Cho}, {Christian}, {Conroy}, {Conway}, {Cordes},
  {Crawford}, {Crew}, {Cruz-Osorio}, {Cui}, {Davelaar}, {De Laurentis},
  {Deane}, {Dempsey}, {Desvignes}, {Dexter}, {Dhruv}, {Doeleman}, {Dougal},
  {Dzib}, {Eatough}, {Emami}, {Falcke}, {Farah}, {Fish}, {Fomalont}, {Ford},
  {Fraga-Encinas}, {Freeman}, {Friberg}, {Fromm}, {Fuentes}, {Galison},
  {Gammie}, {Garc{\'\i}a}, {Gentaz}, {Georgiev}, {Goddi}, {Gold},
  {G{\'o}mez-Ruiz}, {G{\'o}mez}, {Gu}, {Gurwell}, {Hada}, {Haggard}, {Haworth},
  {Hecht}, {Hesper}, {Heumann}, {Ho}, {Ho}, {Honma}, {Huang}, {Huang},
  {Hughes}, {Ikeda}, {Violette Impellizzeri}, {Inoue}, {Issaoun}, {James},
  {Jannuzi}, {Janssen}, {Jeter}, {Jiang}, {Jim{\'e}nez-Rosales}, {Johnson},
  {Jorstad}, {Joshi}, {Jung}, {Karami}, {Karuppusamy}, {Kawashima}, {Keating},
  {Kettenis}, {Kim}, {Kim}, {Kim}, {Kim}, {Kino}, {Koay}, {Kocherlakota},
  {Kofuji}, {Koch}, {Koyama}, {Kramer}, {Kramer}, {Krichbaum}, {Kuo}, {Bella},
  {Lauer}, {Lee}, {Lee}, {Leung}, {Levis}, {Li}, {Lico}, {Lindahl},
  {Lindqvist}, {Lisakov}, {Liu}, {Liu}, {Liuzzo}, {Lo}, {Lobanov}, {Loinard},
  {Lonsdale}, {Lu}, {Mao}, {Marchili}, {Markoff}, {Marrone}, {Marscher},
  {Mart{\'\i}-Vidal}, {Matsushita}, {Matthews}, {Medeiros}, {Menten},
  {Michalik}, {Mizuno}, {Mizuno}, {Moran}, {Moriyama}, {Moscibrodzka},
  {M{\"u}ller}, {Mus}, {Musoke}, {Myserlis}, {Nadolski}, {Nagai}, {Nagar},
  {Nakamura}, {Narayan}, {Narayanan}, {Natarajan}, {Nathanail}, {Navarro
  Fuentes}, {Neilsen}, {Neri}, {Ni}, {Noutsos}, {Nowak}, {Oh}, {Okino},
  {Olivares}, {Ortiz-Le{\'o}n}, {Oyama}, {{\"O}zel}, {Palumbo}, {Filippos
  Paraschos}, {Park}, {Parsons}, {Patel}, {Pen}, {Pesce}, {Pi{\'e}tu},
  {Plambeck}, {PopStefanija}, {Porth}, {P{\"o}tzl}, {Prather},
  {Preciado-L{\'o}pez}, {Psaltis}, {Pu}, {Ramakrishnan}, {Rao}, {Rawlings},
  {Raymond}, {Rezzolla}, {Ricarte}, {Ripperda}, {Roelofs}, {Rogers}, {Ros},
  {Romero-Ca{\~n}izales}, {Roshanineshat}, {Rottmann}, {Roy}, {Ruiz},
  {Ruszczyk}, {Rygl}, {S{\'a}nchez}, {S{\'a}nchez-Arg{\"u}elles},
  {S{\'a}nchez-Portal}, {Sasada}, {Satapathy}, {Savolainen}, {Schloerb},
  {Schonfeld}, {Schuster}, {Shao}, {Shen}, {Small}, {Sohn}, {SooHoo},
  {Souccar}, {Sun}, {Tazaki}, {Tetarenko}, {Tiede}, {Tilanus}, {Titus},
  {Torne}, {Traianou}, {Trent}, {Trippe}, {Turk}, {van Bemmel}, {van
  Langevelde}, {van Rossum}, {Vos}, {Wagner}, {Ward-Thompson}, {Wardle},
  {Weintroub}, {Wex}, {Wharton}, {Wielgus}, {Wiik}, {Witzel}, {Wondrak},
  {Wong}, {Wu}, {Yamaguchi}, {Yoon}, {Young}, {Young}, {Younsi}, {Yuan},
  {Yuan}, {Zensus}, {Zhang}, {Zhao}, {Zhao}, {Chan}, {Qiu}, {Ressler}, and
  {White}]{EHT_sgr_5}
{Event Horizon Telescope Collaboration}.; {Akiyama}, K.; {Alberdi}, A.; {Alef},
  W.; {Carlos Algaba}, J.; {Anantua}, R.; {Asada}, K.; {Azulay}, R.; {Bach},
  U.; {Baczko}, A.K.;  et~al.
\newblock {First Sagittarius A* Event Horizon Telescope Results. V. Testing
  Astrophysical Models of the Galactic Center Black Hole}.
\newblock {\em \apjl} {\bf 2022}, {\em 930},~L16.
\newblock {\url{https://doi.org/10.3847/2041-8213/ac6672}}.

\bibitem[{Narayan} \em{et~al.}(2003){Narayan}, {Igumenshchev}, and
  {Abramowicz}]{Narayan_2003}
{Narayan}, R.; {Igumenshchev}, I.V.; {Abramowicz}, M.A.
\newblock {Magnetically Arrested Disk: an Energetically Efficient Accretion
  Flow}.
\newblock {\em \pasj} {\bf 2003}, {\em 55},~L69--L72,
  \href{http://xxx.lanl.gov/abs/astro-ph/0305029}{{\normalfont
  [arXiv:astro-ph/astro-ph/0305029]}}.
\newblock {\url{https://doi.org/10.1093/pasj/55.6.L69}}.

\bibitem[{Igumenshchev} \em{et~al.}(2003){Igumenshchev}, {Narayan}, and
  {Abramowicz}]{Igumenshchev_2003}
{Igumenshchev}, I.V.; {Narayan}, R.; {Abramowicz}, M.A.
\newblock {Three-dimensional Magnetohydrodynamic Simulations of Radiatively
  Inefficient Accretion Flows}.
\newblock {\em \apj} {\bf 2003}, {\em 592},~1042--1059,
  \href{http://xxx.lanl.gov/abs/astro-ph/0301402}{{\normalfont
  [arXiv:astro-ph/astro-ph/0301402]}}.
\newblock {\url{https://doi.org/10.1086/375769}}.

\bibitem[{Ripperda} \em{et~al.}(2022){Ripperda}, {Liska}, {Chatterjee},
  {Musoke}, {Philippov}, {Markoff}, {Tchekhovskoy}, and
  {Younsi}]{Ripperda_2022}
{Ripperda}, B.; {Liska}, M.; {Chatterjee}, K.; {Musoke}, G.; {Philippov}, A.A.;
  {Markoff}, S.B.; {Tchekhovskoy}, A.; {Younsi}, Z.
\newblock {Black Hole Flares: Ejection of Accreted Magnetic Flux through 3D
  Plasmoid-mediated Reconnection}.
\newblock {\em \apjl} {\bf 2022}, {\em 924},~L32,
  \href{http://xxx.lanl.gov/abs/2109.15115}{{\normalfont
  [arXiv:astro-ph.HE/2109.15115]}}.
\newblock {\url{https://doi.org/10.3847/2041-8213/ac46a1}}.

\bibitem[{Wielgus} \em{et~al.}(2022){Wielgus}, {Moscibrodzka}, {Vos}, {Gelles},
  {Mart{\'\i}-Vidal}, {Farah}, {Marchili}, {Goddi}, and
  {Messias}]{Wielgus:2022b}
{Wielgus}, M.; {Moscibrodzka}, M.; {Vos}, J.; {Gelles}, Z.; {Mart{\'\i}-Vidal},
  I.; {Farah}, J.; {Marchili}, N.; {Goddi}, C.; {Messias}, H.
\newblock {Orbital motion near Sagittarius A$^{*}$ . Constraints from
  polarimetric ALMA observations}.
\newblock {\em \aap} {\bf 2022}, {\em 665},~L6,
  \href{http://xxx.lanl.gov/abs/2209.09926}{{\normalfont
  [arXiv:astro-ph.HE/2209.09926]}}.
\newblock {\url{https://doi.org/10.1051/0004-6361/202244493}}.

\bibitem[{Narayan} \em{et~al.}(2022){Narayan}, {Chael}, {Chatterjee},
  {Ricarte}, and {Curd}]{Narayan_2022}
{Narayan}, R.; {Chael}, A.; {Chatterjee}, K.; {Ricarte}, A.; {Curd}, B.
\newblock {Jets in magnetically arrested hot accretion flows: geometry, power,
  and black hole spin-down}.
\newblock {\em \mnras} {\bf 2022}, {\em 511},~3795--3813,
  \href{http://xxx.lanl.gov/abs/2108.12380}{{\normalfont
  [arXiv:astro-ph.HE/2108.12380]}}.
\newblock {\url{https://doi.org/10.1093/mnras/stac285}}.

\bibitem[{Event Horizon Telescope Collaboration} \em{et~al.}(2022){Event
  Horizon Telescope Collaboration}, {Akiyama}, {Alberdi}, {Alef}, {Algaba},
  {Anantua}, {Asada}, {Azulay}, {Bach}, {Baczko}, {Ball}, {Balokovi{\'c}},
  {Barrett}, {Baub{\"o}ck}, {Benson}, {Bintley}, {Blackburn}, {Blundell},
  {Bouman}, {Bower}, {Boyce}, {Bremer}, {Brinkerink}, {Brissenden}, {Britzen},
  {Broderick}, {Broguiere}, {Bronzwaer}, {Bustamante}, {Byun}, {Carlstrom},
  {Ceccobello}, {Chael}, {Chan}, {Chatterjee}, {Chatterjee}, {Chen}, {Chen},
  {Cheng}, {Cho}, {Christian}, {Conroy}, {Conway}, {Cordes}, {Crawford},
  {Crew}, {Cruz-Osorio}, {Cui}, {Davelaar}, {De Laurentis}, {Deane}, {Dempsey},
  {Desvignes}, {Dexter}, {Dhruv}, {Doeleman}, {Dougal}, {Dzib}, {Eatough},
  {Emami}, {Falcke}, {Farah}, {Fish}, {Fomalont}, {Ford}, {Fraga-Encinas},
  {Freeman}, {Friberg}, {Fromm}, {Fuentes}, {Galison}, {Gammie}, {Garc{\'\i}a},
  {Gentaz}, {Georgiev}, {Goddi}, {Gold}, {G{\'o}mez-Ruiz}, {G{\'o}mez}, {Gu},
  {Gurwell}, {Hada}, {Haggard}, {Haworth}, {Hecht}, {Hesper}, {Heumann}, {Ho},
  {Ho}, {Honma}, {Huang}, {Huang}, {Hughes}, {Ikeda}, {Impellizzeri}, {Inoue},
  {Issaoun}, {James}, {Jannuzi}, {Janssen}, {Jeter}, {Jiang},
  {Jim{\'e}nez-Rosales}, {Johnson}, {Jorstad}, {Joshi}, {Jung}, {Karami},
  {Karuppusamy}, {Kawashima}, {Keating}, {Kettenis}, {Kim}, {Kim}, {Kim},
  {Kim}, {Kino}, {Koay}, {Kocherlakota}, {Kofuji}, {Koch}, {Koyama}, {Kramer},
  {Kramer}, {Krichbaum}, {Kuo}, {Bella}, {Lauer}, {Lee}, {Lee}, {Leung},
  {Levis}, {Li}, {Lico}, {Lindahl}, {Lindqvist}, {Lisakov}, {Liu}, {Liu},
  {Liuzzo}, {Lo}, {Lobanov}, {Loinard}, {Lonsdale}, {Lu}, {Mao}, {Marchili},
  {Markoff}, {Marrone}, {Marscher}, {Mart{\'\i}-Vidal}, {Matsushita},
  {Matthews}, {Medeiros}, {Menten}, {Michalik}, {Mizuno}, {Mizuno}, {Moran},
  {Moriyama}, {Moscibrodzka}, {M{\"u}ller}, {Mus}, {Musoke}, {Myserlis},
  {Nadolski}, {Nagai}, {Nagar}, {Nakamura}, {Narayan}, {Narayanan},
  {Natarajan}, {Nathanail}, {Fuentes}, {Neilsen}, {Neri}, {Ni}, {Noutsos},
  {Nowak}, {Oh}, {Okino}, {Olivares}, {Ortiz-Le{\'o}n}, {Oyama}, {{\"O}zel},
  {Palumbo}, {Paraschos}, {Park}, {Parsons}, {Patel}, {Pen}, {Pesce},
  {Pi{\'e}tu}, {Plambeck}, {PopStefanija}, {Porth}, {P{\"o}tzl}, {Prather},
  {Preciado-L{\'o}pez}, {Psaltis}, {Pu}, {Ramakrishnan}, {Rao}, {Rawlings},
  {Raymond}, {Rezzolla}, {Ricarte}, {Ripperda}, {Roelofs}, {Rogers}, {Ros},
  {Romero-Ca{\~n}izales}, {Roshanineshat}, {Rottmann}, {Roy}, {Ruiz},
  {Ruszczyk}, {Rygl}, {S{\'a}nchez}, {S{\'a}nchez-Arg{\"u}elles},
  {S{\'a}nchez-Portal}, {Sasada}, {Satapathy}, {Savolainen}, {Schloerb},
  {Schonfeld}, {Schuster}, {Shao}, {Shen}, {Small}, {Sohn}, {SooHoo},
  {Souccar}, {Sun}, {Tazaki}, {Tetarenko}, {Tiede}, {Tilanus}, {Titus},
  {Torne}, {Traianou}, {Trent}, {Trippe}, {Turk}, {van Bemmel}, {van
  Langevelde}, {van Rossum}, {Vos}, {Wagner}, {Ward-Thompson}, {Wardle},
  {Weintroub}, {Wex}, {Wharton}, {Wielgus}, {Wiik}, {Witzel}, {Wondrak},
  {Wong}, {Wu}, {Yamaguchi}, {Yoon}, {Young}, {Young}, {Younsi}, {Yuan},
  {Yuan}, {Zensus}, {Zhang}, {Zhao}, {Zhao}, {Agurto}, {Araneda}, {Arriagada},
  {Bertarini}, {Berthold}, {Blanchard}, {Brown}, {C{\'a}rdenas}, {Cantzler},
  {Caro}, {Chuter}, {Ciechanowicz}, {Coulson}, {Crowley}, {Degenaar},
  {Dornbusch}, {Dur{\'a}n}, {Forster}, {Geertsema}, {Gonz{\'a}lez}, {Graham},
  {Gueth}, {Han}, {Herrera}, {Herrero-Illana}, {Heyminck}, {Hoge}, {Huang},
  {Jiang}, {John}, {Klein}, {Kubo}, {Kuroda}, {Kwon}, {Laing}, {Liu}, {Liu},
  {Mac-Auliffe}, {Martin-Cocher}, {Matulonis}, {Messias}, {Meyer-Zhao},
  {Montenegro-Montes}, {Montgomerie}, {Muders}, {Nishioka}, {Norton},
  {Olivares}, {P{\'e}rez-Beaupuits}, {Parra}, {Poirier}, {Pradel}, {Raffin},
  {Ram{\'\i}rez}, {Reynolds}, {Saez-Madain}, {Santana}, {Silva}, {Sousa},
  {Stahm}, {Torstensson}, {Venegas}, {Walther}, {Wieching}, {Wijnands}, and
  {Wouterloot}]{EHT_sgr_2}
{Event Horizon Telescope Collaboration}.; {Akiyama}, K.; {Alberdi}, A.; {Alef},
  W.; {Algaba}, J.C.; {Anantua}, R.; {Asada}, K.; {Azulay}, R.; {Bach}, U.;
  {Baczko}, A.K.;  et~al.
\newblock {First Sagittarius A* Event Horizon Telescope Results. II. EHT and
  Multiwavelength Observations, Data Processing, and Calibration}.
\newblock {\em \apjl} {\bf 2022}, {\em 930},~L13.
\newblock {\url{https://doi.org/10.3847/2041-8213/ac6675}}.

\bibitem[{Wielgus} \em{et~al.}(2022){Wielgus}, {Marchili}, {Mart{\'\i}-Vidal},
  {Keating}, {Ramakrishnan}, {Tiede}, {Fomalont}, {Issaoun}, {Neilsen},
  {Nowak}, {Blackburn}, {Gammie}, {Goddi}, {Haggard}, {Lee}, {Moscibrodzka},
  {Tetarenko}, {Bower}, {Chan}, {Chatterjee}, {Chesler}, {Dexter}, {Doeleman},
  {Georgiev}, {Gurwell}, {Johnson}, {Marrone}, {Mus}, {Psaltis}, {Ripperda},
  {Witzel}, {Akiyama}, {Alberdi}, {Alef}, {Algaba}, {Anantua}, {Asada},
  {Azulay}, {Bach}, {Baczko}, {Ball}, {Balokovi{\'c}}, {Barrett},
  {Baub{\"o}ck}, {Benson}, {Bintley}, {Blundell}, {Boland}, {Bouman}, {Boyce},
  {Bremer}, {Brinkerink}, {Brissenden}, {Britzen}, {Broderick}, {Broguiere},
  {Bronzwaer}, {Bustamante}, {Byun}, {Carlstrom}, {Ceccobello}, {Chael},
  {Chatterjee}, {Chen}, {Chen}, {Cho}, {Christian}, {Conroy}, {Conway},
  {Cordes}, {Crawford}, {Crew}, {Cruz-Osorio}, {Cui}, {Davelaar}, {De
  Laurentis}, {Deane}, {Dempsey}, {Desvignes}, {Dhruv}, {Dzib}, {Eatough},
  {Emami}, {Falcke}, {Farah}, {Fish}, {Ford}, {Fraga-Encinas}, {Freeman},
  {Friberg}, {Fromm}, {Fuentes}, {Galison}, {Garc{\'\i}a}, {Gentaz}, {Gold},
  {G{\'o}mez-Ruiz}, {G{\'o}mez}, {Gu}, {Hada}, {Haworth}, {Hecht}, {Hesper},
  {Ho}, {Ho}, {Honma}, {Huang}, {Huang}, {Hughes}, {Ikeda}, {Impellizzeri},
  {Inoue}, {James}, {Jannuzi}, {Janssen}, {Jeter}, {Jiang},
  {Jim{\'e}nez-Rosales}, {Jorstad}, {Joshi}, {Jung}, {Karami}, {Karuppusamy},
  {Kawashima}, {Kettenis}, {Kim}, {Kim}, {Kim}, {Kim}, {Kino}, {Koay},
  {Kocherlakota}, {Kofuji}, {Koch}, {Koyama}, {Kramer}, {Kramer}, {Krichbaum},
  {Kuo}, {La Bella}, {Lauer}, {Lee}, {Leung}, {Levis}, {Li}, {Lico}, {Lindahl},
  {Lindqvist}, {Lisakov}, {Liu}, {Liu}, {Liuzzo}, {Lo}, {Lobanov}, {Loinard},
  {Lonsdale}, {Lu}, {Mao}, {Markoff}, {Marscher}, {Matsushita}, {Matthews},
  {Medeiros}, {Menten}, {Michalik}, {Mizuno}, {Mizuno}, {Moran}, {Moriyama},
  {M{\"u}ller}, {Musoke}, {Myserlis}, {Nadolski}, {Nagai}, {Nagar}, {Nakamura},
  {Narayan}, {Narayanan}, {Natarajan}, {Nathanail}, {Navarro Fuentes}, {Neri},
  {Ni}, {Noutsos}, {Oh}, {Okino}, {Olivares}, {Ortiz-Le{\'o}n}, {Oyama},
  {{\"O}zel}, {Palumbo}, {Paraschos}, {Park}, {Parsons}, {Patel}, {Pen},
  {Pesce}, {Pi{\'e}tu}, {Plambeck}, {PopStefanija}, {Porth}, {P{\"o}tzl},
  {Prather}, {Preciado-L{\'o}pez}, {Pu}, {Rao}, {Rawlings}, {Raymond},
  {Rezzolla}, {Ricarte}, {Roelofs}, {Rogers}, {Ros}, {Romero-Canizales},
  {Roshanineshat}, {Rottmann}, {Roy}, {Ruiz}, {Ruszczyk}, {Rygl},
  {S{\'a}nchez}, {S{\'a}nchez-Arg{\"u}elles}, {S{\'a}nchez-Portal}, {Sasada},
  {Satapathy}, {Savolainen}, {Schloerb}, {Schuster}, {Shao}, {Shen}, {Small},
  {Won Sohn}, {SooHoo}, {Souccar}, {Sun}, {Tazaki}, {Tilanus}, {Titus},
  {Torne}, {Traianou}, {Trent}, {Trippe}, {van Bemmel}, {van Langevelde}, {van
  Rossum}, {Vos}, {Wagner}, {Ward-Thompson}, {Wardle}, {Weintroub}, {Wex},
  {Wharton}, {Wiik}, {Wondrak}, {Wong}, {Wu}, {Yamaguchi}, {Yoon}, {Young},
  {Young}, {Younsi}, {Yuan}, {Yuan}, {Zensus}, {Zhang}, {Zhao}, and
  {Zhao}]{Wielgus:2022a}
{Wielgus}, M.; {Marchili}, N.; {Mart{\'\i}-Vidal}, I.; {Keating}, G.K.;
  {Ramakrishnan}, V.; {Tiede}, P.; {Fomalont}, E.; {Issaoun}, S.; {Neilsen},
  J.; {Nowak}, M.A.;  et~al.
\newblock {Millimeter Light Curves of Sagittarius A* Observed during the 2017
  Event Horizon Telescope Campaign}.
\newblock {\em \apjl} {\bf 2022}, {\em 930},~L19,
  \href{http://xxx.lanl.gov/abs/2207.06829}{{\normalfont
  [arXiv:astro-ph.HE/2207.06829]}}.
\newblock {\url{https://doi.org/10.3847/2041-8213/ac6428}}.

\bibitem[{Haggard} \em{et~al.}(2019){Haggard}, {Nynka}, {Mon}, {de la Cruz
  Hernandez}, {Nowak}, {Heinke}, {Neilsen}, {Dexter}, {Fragile}, {Baganoff},
  {Bower}, {Corrales}, {Coti Zelati}, {Degenaar}, {Markoff}, {Morris}, {Ponti},
  {Rea}, {Wilms}, and {Yusef-Zadeh}]{Haggard:2019}
{Haggard}, D.; {Nynka}, M.; {Mon}, B.; {de la Cruz Hernandez}, N.; {Nowak}, M.;
  {Heinke}, C.; {Neilsen}, J.; {Dexter}, J.; {Fragile}, P.C.; {Baganoff}, F.;
  et~al.
\newblock {Chandra Spectral and Timing Analysis of Sgr A*'s Brightest X-Ray
  Flares}.
\newblock {\em \apj} {\bf 2019}, {\em 886},~96,
  \href{http://xxx.lanl.gov/abs/1908.01781}{{\normalfont
  [arXiv:astro-ph.HE/1908.01781]}}.
\newblock {\url{https://doi.org/10.3847/1538-4357/ab4a7f}}.

\bibitem[{GRAVITY Collaboration} \em{et~al.}(2021){GRAVITY Collaboration},
  {Abuter}, {Amorim}, {Baub{\"o}ck}, {Baganoff}, {Berger}, {Boyce}, {Bonnet},
  {Brandner}, {Cl{\'e}net}, {Davies}, {de Zeeuw}, {Dexter}, {Dallilar},
  {Drescher}, {Eckart}, {Eisenhauer}, {Fazio}, {F{\"o}rster Schreiber},
  {Foster}, {Gammie}, {Garcia}, {Gao}, {Gendron}, {Genzel}, {Ghisellini},
  {Gillessen}, {Gurwell}, {Habibi}, {Haggard}, {Hailey}, {Harrison}, {Haubois},
  {Hei{\ss}el}, {Henning}, {Hippler}, {Hora}, {Horrobin},
  {Jim{\'e}nez-Rosales}, {Jochum}, {Jocou}, {Kaufer}, {Kervella}, {Lacour},
  {Lapeyr{\`e}re}, {Le Bouquin}, {L{\'e}na}, {Lowrance}, {Lutz}, {Markoff},
  {Mori}, {Morris}, {Neilsen}, {Nowak}, {Ott}, {Paumard}, {Perraut}, {Perrin},
  {Ponti}, {Pfuhl}, {Rabien}, {Rodr{\'\i}guez-Coira}, {Shangguan}, {Shimizu},
  {Scheithauer}, {Smith}, {Stadler}, {Stern}, {Straub}, {Straubmeier}, {Sturm},
  {Tacconi}, {Vincent}, {von Fellenberg}, {Waisberg}, {Widmann}, {Wieprecht},
  {Wiezorrek}, {Willner}, {Witzel}, {Woillez}, {Yazici}, {Young}, {Zhang}, and
  {Zins}]{GRAVITY:2021}
{GRAVITY Collaboration}.; {Abuter}, R.; {Amorim}, A.; {Baub{\"o}ck}, M.;
  {Baganoff}, F.; {Berger}, J.P.; {Boyce}, H.; {Bonnet}, H.; {Brandner}, W.;
  {Cl{\'e}net}, Y.;  et~al.
\newblock {Constraining particle acceleration in Sgr A$^{{\ensuremath{\star}}}$
  with simultaneous GRAVITY, Spitzer, NuSTAR, and Chandra observations}.
\newblock {\em \aap} {\bf 2021}, {\em 654},~A22,
  \href{http://xxx.lanl.gov/abs/2107.01096}{{\normalfont
  [arXiv:astro-ph.HE/2107.01096]}}.
\newblock {\url{https://doi.org/10.1051/0004-6361/202140981}}.

\bibitem[{Gelles} \em{et~al.}(2021){Gelles}, {Himwich}, {Johnson}, and
  {Palumbo}]{Gelles_2021b}
{Gelles}, Z.; {Himwich}, E.; {Johnson}, M.D.; {Palumbo}, D.C.M.
\newblock {Polarized image of equatorial emission in the Kerr geometry}.
\newblock {\em \prd} {\bf 2021}, {\em 104},~044060,
  \href{http://xxx.lanl.gov/abs/2105.09440}{{\normalfont
  [arXiv:gr-qc/2105.09440]}}.
\newblock {\url{https://doi.org/10.1103/PhysRevD.104.044060}}.

\bibitem[{Mo{\'s}cibrodzka} and {Gammie}(2018)]{ipole_2018}
{Mo{\'s}cibrodzka}, M.; {Gammie}, C.F.
\newblock {IPOLE - semi-analytic scheme for relativistic polarized radiative
  transport}.
\newblock {\em \mnras} {\bf 2018}, {\em 475},~43--54,
  \href{http://xxx.lanl.gov/abs/1712.03057}{{\normalfont
  [arXiv:astro-ph.HE/1712.03057]}}.
\newblock {\url{https://doi.org/10.1093/mnras/stx3162}}.

\bibitem[{Gelles} \em{et~al.}(2021){Gelles}, {Prather}, {Palumbo}, {Johnson},
  {Wong}, and {Georgiev}]{Gelles_2021a}
{Gelles}, Z.; {Prather}, B.S.; {Palumbo}, D.C.M.; {Johnson}, M.D.; {Wong},
  G.N.; {Georgiev}, B.
\newblock {The Role of Adaptive Ray Tracing in Analyzing Black Hole Structure}.
\newblock {\em \apj} {\bf 2021}, {\em 912},~39,
  \href{http://xxx.lanl.gov/abs/2103.07417}{{\normalfont
  [arXiv:astro-ph.HE/2103.07417]}}.
\newblock {\url{https://doi.org/10.3847/1538-4357/abee13}}.

\bibitem[Liska \em{et~al.}(2019)Liska, Chatterjee, Tchekhovskoy, Yoon, van
  Eijnatten, Hesp, Markoff, Ingram, and van~der Klis]{Liska_hamr}
Liska, M.; Chatterjee, K.; Tchekhovskoy, A.; Yoon, D.; van Eijnatten, D.; Hesp,
  C.; Markoff, S.; Ingram, A.; van~der Klis, M.
\newblock H-AMR: A New GPU-accelerated GRMHD Code for Exascale Computing With
  3D Adaptive Mesh Refinement and Local Adaptive Time-stepping,  2019,
  \href{http://xxx.lanl.gov/abs/1912.10192}{{\normalfont
  [arXiv:astro-ph.HE/1912.10192]}}.

\bibitem[{Narayan} \em{et~al.}(2021){Narayan}, {Palumbo}, {Johnson}, {Gelles},
  {Himwich}, {Chang}, {Ricarte}, {Dexter}, {Gammie}, {Chael}, {Event Horizon
  Telescope Collaboration}, {Akiyama}, {Alberdi}, {Alef}, {Algaba}, {Anantua},
  {Asada}, {Azulay}, {Baczko}, {Ball}, {Balokovi{\'c}}, {Barrett}, {Benson},
  {Bintley}, {Blackburn}, {Blundell}, {Boland}, {Bouman}, {Bower}, {Boyce},
  {Bremer}, {Brinkerink}, {Brissenden}, {Britzen}, {Broderick}, {Broguiere},
  {Bronzwaer}, {Byun}, {Carlstrom}, {Chan}, {Chatterjee}, {Chatterjee}, {Chen},
  {Chen}, {Chesler}, {Cho}, {Christian}, {Conway}, {Cordes}, {Crawford},
  {Crew}, {Cruz-Osorio}, {Cui}, {Davelaar}, {De Laurentis}, {Deane}, {Dempsey},
  {Desvignes}, {Doeleman}, {Eatough}, {Falcke}, {Farah}, {Fish}, {Fomalont},
  {Ford}, {Fraga-Encinas}, {Friberg}, {Fromm}, {Fuentes}, {Galison},
  {Garc{\'\i}a}, {Gentaz}, {Georgiev}, {Goddi}, {Gold}, {G{\'o}mez},
  {G{\'o}mez-Ruiz}, {Gu}, {Gurwell}, {Hada}, {Haggard}, {Hecht}, {Hesper},
  {Ho}, {Ho}, {Honma}, {Huang}, {Huang}, {Hughes}, {Ikeda}, {Inoue}, {Issaoun},
  {James}, {Jannuzi}, {Janssen}, {Jeter}, {Jiang}, {Jimenez-Rosales},
  {Jorstad}, {Jung}, {Karami}, {Karuppusamy}, {Kawashima}, {Keating},
  {Kettenis}, {Kim}, {Kim}, {Kim}, {Kim}, {Kino}, {Koay}, {Kofuji}, {Koch},
  {Koyama}, {Kramer}, {Kramer}, {Krichbaum}, {Kuo}, {Lauer}, {Lee}, {Levis},
  {Li}, {Li}, {Lindqvist}, {Lico}, {Lindahl}, {Liu}, {Liu}, {Liuzzo}, {Lo},
  {Lobanov}, {Loinard}, {Lonsdale}, {Lu}, {MacDonald}, {Mao}, {Marchili},
  {Markoff}, {Marrone}, {Marscher}, {Mart{\'\i}-Vidal}, {Matsushita},
  {Matthews}, {Medeiros}, {Menten}, {Mizuno}, {Mizuno}, {Moran}, {Moriyama},
  {Moscibrodzka}, {M{\"u}ller}, {Musoke}, {Mej{\'\i}as}, {Nagai}, {Nagar},
  {Nakamura}, {Narayanan}, {Natarajan}, {Nathanail}, {Neilsen}, {Neri}, {Ni},
  {Noutsos}, {Nowak}, {Okino}, {Olivares}, {Ortiz-Le{\'o}n}, {Oyama},
  {{\"O}zel}, {Park}, {Patel}, {Pen}, {Pesce}, {Pi{\'e}tu}, {Plambeck},
  {PopStefanija}, {Porth}, {P{\"o}tzl}, {Prather}, {Preciado-L{\'o}pez},
  {Psaltis}, {Pu}, {Ramakrishnan}, {Rao}, {Rawlings}, {Raymond}, {Rezzolla},
  {Ripperda}, {Roelofs}, {Rogers}, {Ros}, {Rose}, {Roshanineshat}, {Rottmann},
  {Roy}, {Ruszczyk}, {Rygl}, {S{\'a}nchez}, {S{\'a}nchez-Arguelles}, {Sasada},
  {Savolainen}, {Schloerb}, {Schuster}, {Shao}, {Shen}, {Small}, {Sohn},
  {SooHoo}, {Sun}, {Tazaki}, {Tetarenko}, {Tiede}, {Tilanus}, {Titus}, {Toma},
  {Torne}, {Trent}, {Traianou}, {Trippe}, {van Bemmel}, {van Langevelde}, {van
  Rossum}, {Wagner}, {Ward-Thompson}, {Wardle}, {Weintroub}, {Wex}, {Wharton},
  {Wielgus}, {Wong}, {Wu}, {Yoon}, {Young}, {Young}, {Younsi}, {Yuan}, {Yuan},
  {Zensus}, {Zhao}, and {Zhao}]{Narayan_2021}
{Narayan}, R.; {Palumbo}, D.C.M.; {Johnson}, M.D.; {Gelles}, Z.; {Himwich}, E.;
  {Chang}, D.O.; {Ricarte}, A.; {Dexter}, J.; {Gammie}, C.F.; {Chael}, A.A.;
  et~al.
\newblock {The Polarized Image of a Synchrotron-emitting Ring of Gas Orbiting a
  Black Hole}.
\newblock {\em \apj} {\bf 2021}, {\em 912},~35,
  \href{http://xxx.lanl.gov/abs/2105.01804}{{\normalfont
  [arXiv:astro-ph.HE/2105.01804]}}.
\newblock {\url{https://doi.org/10.3847/1538-4357/abf117}}.

\bibitem[{Event Horizon Telescope Collaboration} \em{et~al.}(2021){Event
  Horizon Telescope Collaboration}, {Akiyama}, {Algaba}, {Alberdi}, {Alef},
  {Anantua}, {Asada}, {Azulay}, {Baczko}, {Ball}, {Balokovi{\'c}}, {Barrett},
  {Benson}, {Bintley}, {Blackburn}, {Blundell}, {Boland}, {Bouman}, {Bower},
  {Boyce}, {Bremer}, {Brinkerink}, {Brissenden}, {Britzen}, {Broderick},
  {Broguiere}, {Bronzwaer}, {Byun}, {Carlstrom}, {Chael}, {Chan}, {Chatterjee},
  {Chatterjee}, {Chen}, {Chen}, {Chesler}, {Cho}, {Christian}, {Conway},
  {Cordes}, {Crawford}, {Crew}, {Cruz-Osorio}, {Cui}, {Davelaar}, {De
  Laurentis}, {Deane}, {Dempsey}, {Desvignes}, {Dexter}, {Doeleman}, {Eatough},
  {Falcke}, {Farah}, {Fish}, {Fomalont}, {Ford}, {Fraga-Encinas}, {Friberg},
  {Fromm}, {Fuentes}, {Galison}, {Gammie}, {Garc{\'\i}a}, {Gelles}, {Gentaz},
  {Georgiev}, {Goddi}, {Gold}, {G{\'o}mez}, {G{\'o}mez-Ruiz}, {Gu}, {Gurwell},
  {Hada}, {Haggard}, {Hecht}, {Hesper}, {Himwich}, {Ho}, {Ho}, {Honma},
  {Huang}, {Huang}, {Hughes}, {Ikeda}, {Inoue}, {Issaoun}, {James}, {Jannuzi},
  {Janssen}, {Jeter}, {Jiang}, {Jimenez-Rosales}, {Johnson}, {Jorstad}, {Jung},
  {Karami}, {Karuppusamy}, {Kawashima}, {Keating}, {Kettenis}, {Kim}, {Kim},
  {Kim}, {Kim}, {Kino}, {Koay}, {Kofuji}, {Koch}, {Koyama}, {Kramer}, {Kramer},
  {Krichbaum}, {Kuo}, {Lauer}, {Lee}, {Levis}, {Li}, {Li}, {Lindqvist}, {Lico},
  {Lindahl}, {Liu}, {Liu}, {Liuzzo}, {Lo}, {Lobanov}, {Loinard}, {Lonsdale},
  {Lu}, {MacDonald}, {Mao}, {Marchili}, {Markoff}, {Marrone}, {Marscher},
  {Mart{\'\i}-Vidal}, {Matsushita}, {Matthews}, {Medeiros}, {Menten}, {Mizuno},
  {Mizuno}, {Moran}, {Moriyama}, {Moscibrodzka}, {M{\"u}ller}, {Musoke}, {Mus
  Mej{\'\i}as}, {Michalik}, {Nadolski}, {Nagai}, {Nagar}, {Nakamura},
  {Narayan}, {Narayanan}, {Natarajan}, {Nathanail}, {Neilsen}, {Neri}, {Ni},
  {Noutsos}, {Nowak}, {Okino}, {Olivares}, {Ortiz-Le{\'o}n}, {Oyama},
  {{\"O}zel}, {Palumbo}, {Park}, {Patel}, {Pen}, {Pesce}, {Pi{\'e}tu},
  {Plambeck}, {PopStefanija}, {Porth}, {P{\"o}tzl}, {Prather},
  {Preciado-L{\'o}pez}, {Psaltis}, {Pu}, {Ramakrishnan}, {Rao}, {Rawlings},
  {Raymond}, {Rezzolla}, {Ricarte}, {Ripperda}, {Roelofs}, {Rogers}, {Ros},
  {Rose}, {Roshanineshat}, {Rottmann}, {Roy}, {Ruszczyk}, {Rygl},
  {S{\'a}nchez}, {S{\'a}nchez-Arguelles}, {Sasada}, {Savolainen}, {Schloerb},
  {Schuster}, {Shao}, {Shen}, {Small}, {Sohn}, {SooHoo}, {Sun}, {Tazaki},
  {Tetarenko}, {Tiede}, {Tilanus}, {Titus}, {Toma}, {Torne}, {Trent},
  {Traianou}, {Trippe}, {van Bemmel}, {van Langevelde}, {van Rossum}, {Wagner},
  {Ward-Thompson}, {Wardle}, {Weintroub}, {Wex}, {Wharton}, {Wielgus}, {Wong},
  {Wu}, {Yoon}, {Young}, {Young}, {Younsi}, {Yuan}, {Yuan}, {Zensus}, {Zhao},
  and {Zhao}]{EHT_8}
{Event Horizon Telescope Collaboration}.; {Akiyama}, K.; {Algaba}, J.C.;
  {Alberdi}, A.; {Alef}, W.; {Anantua}, R.; {Asada}, K.; {Azulay}, R.;
  {Baczko}, A.K.; {Ball}, D.;  et~al.
\newblock {First M87 Event Horizon Telescope Results. VIII. Magnetic Field
  Structure near The Event Horizon}.
\newblock {\em \apjl} {\bf 2021}, {\em 910},~L13,
  \href{http://xxx.lanl.gov/abs/2105.01173}{{\normalfont
  [arXiv:astro-ph.HE/2105.01173]}}.
\newblock {\url{https://doi.org/10.3847/2041-8213/abe4de}}.

\bibitem[{Ricarte} \em{et~al.}(2022){Ricarte}, {Gammie}, {Narayan}, and
  {Prather}]{Ricarte_2022}
{Ricarte}, A.; {Gammie}, C.; {Narayan}, R.; {Prather}, B.S.
\newblock {Probing Plasma Physics with Spectral Index Maps of Accreting Black
  Holes on Event Horizon Scales}.
\newblock {\em arXiv e-prints} {\bf 2022}, p. arXiv:2202.02408,
  \href{http://xxx.lanl.gov/abs/2202.02408}{{\normalfont
  [arXiv:astro-ph.HE/2202.02408]}}.

\bibitem[{Boyce} \em{et~al.}(2022){Boyce}, {Haggard}, {Witzel}, {Fellenberg},
  {Willner}, {Becklin}, {Do}, {Eckart}, {Fazio}, {Gurwell}, {Hora}, {Markoff},
  {Morris}, {Neilsen}, {Nowak}, {Smith}, and {Zhang}]{Boyce:2022}
{Boyce}, H.; {Haggard}, D.; {Witzel}, G.; {Fellenberg}, S.v.; {Willner}, S.P.;
  {Becklin}, E.E.; {Do}, T.; {Eckart}, A.; {Fazio}, G.G.; {Gurwell}, M.A.;
  et~al.
\newblock {Multiwavelength Variability of Sagittarius A* in 2019 July}.
\newblock {\em \apj} {\bf 2022}, {\em 931},~7,
  \href{http://xxx.lanl.gov/abs/2203.13311}{{\normalfont
  [arXiv:astro-ph.HE/2203.13311]}}.
\newblock {\url{https://doi.org/10.3847/1538-4357/ac6104}}.

\bibitem[{Aharonian} \em{et~al.}(2006){Aharonian}, {Akhperjanian},
  {Bazer-Bachi}, {Beilicke}, {Benbow}, {Berge}, {Bernl{\"o}hr}, {Boisson},
  {Bolz}, {Borrel}, {Braun}, {Brown}, {B{\"u}hler}, {B{\"u}sching}, {Carrigan},
  {Chadwick}, {Chounet}, {Coignet}, {Cornils}, {Costamante}, {Degrange},
  {Dickinson}, {Djannati-Ata{\"\i}}, {Drury}, {Dubus}, {Egberts},
  {Emmanoulopoulos}, {Espigat}, {Feinstein}, {Ferrero}, {Fiasson}, {Fontaine},
  {Funk}, {Funk}, {F{\"u}{\ss}ling}, {Gallant}, {Giebels}, {Glicenstein},
  {Goret}, {Hadjichristidis}, {Hauser}, {Hauser}, {Heinzelmann}, {Henri},
  {Hermann}, {Hinton}, {Hoffmann}, {Hofmann}, {Holleran}, {Hoppe}, {Horns},
  {Jacholkowska}, {de Jager}, {Kendziorra}, {Kerschhaggl}, {Kh{\'e}lifi},
  {Komin}, {Konopelko}, {Kosack}, {Lamanna}, {Latham}, {Le Gallou},
  {Lemi{\`e}re}, {Lemoine-Goumard}, {Lenain}, {Lohse}, {Martin},
  {Martineau-Huynh}, {Marcowith}, {Masterson}, {Maurin}, {McComb}, {Moulin},
  {de Naurois}, {Nedbal}, {Nolan}, {Noutsos}, {Orford}, {Osborne}, {Ouchrif},
  {Panter}, {Pelletier}, {Pita}, {P{\"u}hlhofer}, {Punch}, {Ranchon},
  {Raubenheimer}, {Raue}, {Rayner}, {Reimer}, {Ripken}, {Rob}, {Rolland},
  {Rosier-Lees}, {Rowell}, {Sahakian}, {Santangelo}, {Saug{\'e}}, {Schlenker},
  {Schlickeiser}, {Schr{\"o}der}, {Schwanke}, {Schwarzburg}, {Schwemmer},
  {Shalchi}, {Sol}, {Spangler}, {Spanier}, {Steenkamp}, {Stegmann}, {Superina},
  {Tam}, {Tavernet}, {Terrier}, {Tluczykont}, {van Eldik}, {Vasileiadis},
  {Venter}, {Vialle}, {Vincent}, {V{\"o}lk}, {Wagner}, and
  {Ward}]{Aharonian:2006}
{Aharonian}, F.; {Akhperjanian}, A.G.; {Bazer-Bachi}, A.R.; {Beilicke}, M.;
  {Benbow}, W.; {Berge}, D.; {Bernl{\"o}hr}, K.; {Boisson}, C.; {Bolz}, O.;
  {Borrel}, V.;  et~al.
\newblock {Fast Variability of Tera-Electron Volt {\ensuremath{\gamma}} Rays
  from the Radio Galaxy M87}.
\newblock {\em Science} {\bf 2006}, {\em 314},~1424--1427,
  \href{http://xxx.lanl.gov/abs/astro-ph/0612016}{{\normalfont
  [arXiv:astro-ph/astro-ph/0612016]}}.
\newblock {\url{https://doi.org/10.1126/science.1134408}}.

\bibitem[{Broderick} and {Loeb}(2005)]{Broderick_2005}
{Broderick}, A.E.; {Loeb}, A.
\newblock {Imaging bright-spots in the accretion flow near the black hole
  horizon of Sgr A*}.
\newblock {\em \mnras} {\bf 2005}, {\em 363},~353--362,
  \href{http://xxx.lanl.gov/abs/astro-ph/0506433}{{\normalfont
  [arXiv:astro-ph/astro-ph/0506433]}}.
\newblock {\url{https://doi.org/10.1111/j.1365-2966.2005.09458.x}}.

\bibitem[{Broderick} and {Loeb}(2006)]{Broderick_2006}
{Broderick}, A.E.; {Loeb}, A.
\newblock {Imaging optically-thin hotspots near the black hole horizon of Sgr
  A* at radio and near-infrared wavelengths}.
\newblock {\em \mnras} {\bf 2006}, {\em 367},~905--916,
  \href{http://xxx.lanl.gov/abs/astro-ph/0509237}{{\normalfont
  [arXiv:astro-ph/astro-ph/0509237]}}.
\newblock {\url{https://doi.org/10.1111/j.1365-2966.2006.10152.x}}.

\bibitem[{Chael} \em{et~al.}(2016){Chael}, {Johnson}, {Narayan}, {Doeleman},
  {Wardle}, and {Bouman}]{Chael_2016}
{Chael}, A.A.; {Johnson}, M.D.; {Narayan}, R.; {Doeleman}, S.S.; {Wardle},
  J.F.C.; {Bouman}, K.L.
\newblock {High-resolution Linear Polarimetric Imaging for the Event Horizon
  Telescope}.
\newblock {\em \apj} {\bf 2016}, {\em 829},~11,
  \href{http://xxx.lanl.gov/abs/1605.06156}{{\normalfont
  [arXiv:astro-ph.IM/1605.06156]}}.
\newblock {\url{https://doi.org/10.3847/0004-637X/829/1/11}}.

\bibitem[Harris \em{et~al.}(2020)Harris, Millman, van~der Walt, Gommers,
  Virtanen, Cournapeau, Wieser, Taylor, Berg, Smith, Kern, Picus, Hoyer, van
  Kerkwijk, Brett, Haldane, del R{'{\i}}o, Wiebe, Peterson,
  G{'{e}}rard-Marchant, Sheppard, Reddy, Weckesser, Abbasi, Gohlke, and
  Oliphant]{numpy}
Harris, C.R.; Millman, K.J.; van~der Walt, S.J.; Gommers, R.; Virtanen, P.;
  Cournapeau, D.; Wieser, E.; Taylor, J.; Berg, S.; Smith, N.J.;  et~al.
\newblock Array programming with {NumPy}.
\newblock {\em Nature} {\bf 2020}, {\em 585},~357--362.
\newblock {\url{https://doi.org/10.1038/s41586-020-2649-2}}.

\bibitem[Hunter(2007)]{matplotlib}
Hunter, J.D.
\newblock Matplotlib: A 2D graphics environment.
\newblock {\em Computing in Science \& Engineering} {\bf 2007}, {\em
  9},~90--95.
\newblock {\url{https://doi.org/10.1109/MCSE.2007.55}}.

\bibitem[Stanzione \em{et~al.}(2020)Stanzione, West, Evans, Minyard, Ghattas,
  and Panda]{Frontera}
Stanzione, D.; West, J.; Evans, R.T.; Minyard, T.; Ghattas, O.; Panda, D.K.
\newblock Frontera: The Evolution of Leadership Computing at the National
  Science Foundation.
\newblock In Proceedings of the Practice and Experience in Advanced Research
  Computing; Association for Computing Machinery: New York, NY, USA,  2020;
  PEARC '20, p. 106–111.

\bibitem[{Leung} \em{et~al.}(2011){Leung}, {Gammie}, and {Noble}]{Leung:2011}
{Leung}, P.K.; {Gammie}, C.F.; {Noble}, S.C.
\newblock {Numerical Calculation of Magnetobremsstrahlung Emission and
  Absorption Coefficients}.
\newblock {\em \apj} {\bf 2011}, {\em 737},~21.
\newblock {\url{https://doi.org/10.1088/0004-637X/737/1/21}}.

\end{thebibliography}

\end{adjustwidth}
\end{document}